\documentclass[a4paper,fleqn]{cas-dc}
\usepackage[justification=centering]{caption}
\usepackage{bm}
\usepackage[justification=centering]{caption}
\usepackage[numbers]{natbib}
\usepackage{graphicx}
\usepackage{multirow} %%控制表格跨行合并的包
\usepackage{stfloats} %%htbp生效的包, h,here; t, top; b: bottom;, p :current page;
\usepackage[export]{adjustbox}
\usepackage{subfig}
\usepackage{bbding}
\usepackage{amsmath}
\usepackage{amssymb}
\usepackage{booktabs}
\usepackage{hyperref}
\usepackage[misc]{ifsym}
\usepackage{color}
\usepackage{caption}
\usepackage[normalem]{ulem} % 加载 ulem 宏包
\usepackage{xcolor}

\usepackage[inline]{enumitem}
\newcommand{\li}[1]{{\color{red}#1}}

\def\tsc#1{\csdef{#1}{\textsc{\lowercase{#1}}\xspace}}
\tsc{WGM}
\tsc{QE}
\tsc{EP}
\tsc{PMS}
\tsc{BEC}
\tsc{DE}
\begin{document}
\let\WriteBookmarks\relax
\def\floatpagepagefraction{1}
\def\textpagefraction{.001}
\shorttitle{\title{Multi-Granularity Sequence Denoising with Weakly Supervised Signal for Sequential Recommendation}}
\shortauthors{Li et~al.}

%\title[mode = title]{A Mutually Enhanced Multi-Scale Relation-Aware Graph Model  for Argument Pair Extraction}    
%\title[mode = title]{Learning Joint Meta-Network for Cross-domain Recommendation}
%\title[mode = title]{Learning Joint Transfer Meta-Network for Cross-domain Recommendation} 
% \title[mode = title]{Joint Transfer Meta-learning for Cross-domain Recommendation} 
%\title[mode = title]{A Global and Local Graph Mutual Enhancement Based Cascading  Framework  for Multi-Behavior Recommendation (GloME)}

\title[mode = title]{MGSD-WSS: Multi-Granularity Sequence Denoising with Weakly Supervised Signal for Sequential Recommendation} 

% \title[mode = title]\title{A Global and Local Graph Mutual Enhancement Based Cascading  Framework  for Multi-Behavior Recommendation (GloME)}
%\tnotemark[1,2]

%\tnotetext[1]{This document is the results of the research project funded by the National Science Foundation.}

%\tnotetext[2]{The second title footnote which is a longer text matter to fill through the whole text width and overflow into another line in the footnotes area of the first page.}

\author[1]{Liang Li}%[style=chinese]
\ead{liliang@stu.cqut.edu.cn}
\credit{Conceptualization of this study, Methodology, Software, Writing - original draft}
\address[1]{College of Computer Science and Engineering,  Chongqing University of Technology, Chongqing 400054, China}

\author[2]{Zhou Yang}%[style=chinese]
\ead{20031007@fzu.edu.cn}
\credit{Writing - review \& editing, Supervision}
\address[2]{College of Computer and Data Science, Fuzhou University, Fuzhou 350108, China}

\author[1]{Xiaofei Zhu}%[style=chinese]
\ead{zxf@cqut.edu.cn}
\credit{Conceptualization of this study, Methodology, Writing - original draft \& review \& editing, Supervision}
\cormark[1] 
% \address[1]{College of Computer Science and Engineering,  Chongqing University of Technology, Chongqing 400054, China}

\cortext[cor1]{Corresponding author}
%\cortext[cor2]{Principal corresponding author}
%\fntext[fn1]{This is the first author footnote. but is common to third author as well.}
%\fntext[fn2]{Another author footnote, this is a very long footnote and it should be a really long footnote. But this footnote is not yet sufficiently long enough to make two lines of footnote text.}

% \nonumnote{This note has no numbers. In this work we demonstrate $a_b$
%   the formation Y\_1 of a new type of polariton on the interface
%   between a cuprous oxide slab and a polystyrene micro-sphere placed
%   on the slab.
%   }

% \begin{graphicalabstract}
% \includegraphics{figs/grabs.pdf}
% \end{graphicalabstract}

% \begin{highlights}
% \item Research highlights item 1
% \item Research highlights item 2
% \item Research highlights item 3
% \end{highlights}

\begin{abstract}
Sequential recommendation aims to predict the next item based on user interests in historical interaction sequences. Historical interaction sequences often contain irrelevant noisy items,  which significantly hinders the performance of recommendation systems. 
Existing research employs unsupervised methods that indirectly identify item-granularity irrelevant noise by predicting the ground truth item. Since these methods lack explicit noise labels, they are prone to misidentify users' interested items as noise. Additionally, while these methods focus on removing item-granularity noise driven by the ground truth item, they overlook interest-granularity noise, limiting their ability to perform broader denoising based on user interests.
To address these issues, we propose \textbf{M}ulti-\textbf{G}ranularity \textbf{S}equence \textbf{D}enoising with \textbf{W}eakly \textbf{S}upervised \textbf{S}ignal for Sequential Recommendation (MGSD-WSS).
% MGSD-WSS first introduces weakly supervised signals and utilizes the Multiple Gaussian Kernel Perceptron module to accurately identify noisy items in the historical interaction sequence.
MGSD-WSS first introduces the Multiple Gaussian Kernel Perceptron module to map the original and enhance sequence into a common representation space and utilizes weakly supervised signals to accurately identify noisy items in the historical interaction sequence.
Subsequently, it employs the item-granularity denoising module with noise-weighted contrastive learning to obtain denoised item representations.
Then, it extracts target interest representations from the ground truth item and applies noise-weighted contrastive learning to obtain denoised interest representations.
Finally, based on the denoised item and interest representations, MGSD-WSS predicts the next item.
% To address these issues, we propose Multi-Granularity Sequence Denoising with Weakly Supervised Signal for Sequential Recommendation (MGSD-WSS). It first constructs weakly supervised noise signals to train the model for accurate noise identification. Then, it learns denoised sequence representations and employs noise-weighted contrastive learning to enhance item-granularity denoising. Next, it extracts interest representations from the ground truth item and uses noise-weighted contrastive learning to improve interest-granularity denoising.
% Based on denoised sequence representations and interest representations, MGSD-WSS predicts the next item.
Extensive experiments on five datasets demonstrate that the proposed method significantly outperforms state-of-the-art sequence recommendation and denoising models.
Our code is available at https://github.com/lalunex/MGSD-WSS.
\end{abstract}


\maketitle

%%%%%%%%%%%%%%%%%%%%%%%%%%%%%%%%%%%%%%%  introduction start
\section{Introduction}
As a key task in recommendation systems, sequential recommendation aims to predict the next item based on the historical interaction sequence\citep{duan2023long, wu2024personalized, orvieto2023resurrecting}.
In real-world scenarios, historical interaction sequences inevitably contain irrelevant noisy items, such as accidental clicks\cite{tolomei2019you} and malicious fake interactions\cite{zhang2020practical, chen2019data}, which significantly degrade the performance of recommendation systems.

To reduce noise in historical interaction sequences, one line of research applies attention mechanisms\cite{li2020time, luo2020collaborative, yuan2021dual, zhou2023attention, zhu2024multilevel} or filtering algorithms\cite{zhou2022filter, zhu2025preference}, commonly referred to as soft denoising techniques.
Li et al.\cite{li2020time} employ clipped time-aware embeddings in self-attention to filter irrelevant interactions dynamically.
Luo et al.\cite{luo2020collaborative} aggregate neighbor sessions via similarity-based attention weights to dynamically filter noisy interactions and enhance item representations.
Yuan et al.\cite{yuan2021dual} integrate a learnable target embedding with self-attention, adaptively capturing preference signals while suppressing noise.
Zhou et al.\cite{zhou2022filter} apply the Fast Fourier Transform (FFT) to transform sequences into the frequency domain and adaptively filter noise through learnable spectral modulation.
Zhu et al.\cite{zhu2025preference} combine the variational autoencoder with FFT to form a conjugate filter to enhance the preference-related patterns while diminishing high-frequency noise and random noise.
These studies demonstrate promising results; however, since noise items are only mitigated by adjusting their weights instead of being directly removed, sequence modeling continues to be influenced by noise\cite{zhu2024multilevel, zhang2022hierarchical}.

To address this issue, another line of research focuses on generating noise detection signals to explicitly remove noise items\cite{sun2021does, tong2021pattern, chen2022denoising, zhang2022hierarchical, zhu2024multilevel, zhang2024ssdrec}, a process known as hard denoising techniques.
Sun et al.\cite{sun2021does} filter high-loss instances with low uncertainty via Gaussian entropy metrics, directly removing unreliable data from training sets.
Tong et al.\cite{tong2021pattern} utilize pattern-enhanced rewards to guide policy agents in explicitly discarding misaligned sequence items.
Chen et al.\cite{chen2022denoising} prune low-confidence interactions via attention score thresholds, directly eliminating noisy items from sequences.
Zhang et al.\cite{zhang2022hierarchical} adopt a layered denoising approach, leveraging user- and sequence-level signals to remove inconsistent items in historical interaction sequences.
Zhang et al.\cite{zhang2024ssdrec} apply a hierarchical denoising module to gradually refine augmented sequences by filtering out false augmentations and removing all noisy items.

Despite achieving strong results, this research still has the following limitations:
\begin{enumerate*}[label=(\roman*)]
\item The research guides the model to identify potential noisy items using the ground truth item instead of true noise labels. While this unsupervised denoising method yields certain outcomes\li{\cite{wang2025disentangled, deng2023latent}}, it is often less effective than supervised denoising methods with true noise labels in accurately identifying noise\cite{zhu2024multilevel, zhang2022hierarchical}.  Additionally, labeling noise in supervised methods requires significant human resources, making it difficult to implement\cite{yang2023debiased, kang2018self, dang2024repeated}. 
% Therefore, how to accurately identify noisy items with labeled noise signals under low resource consumption is a major challenge.
% \li{Weakly supervised learning can minimize resource consumption while improving noise recognition ability.}
%\zhu{Weakly supervised learning can mitigate the resource consumption as well as maintain a satisfactory noise identification capability.}
% The research guides the model to identify potential noisy items by predicting the ground truth item, enabling unsupervised denoising without true noise labels. However, since there is no direct guidance from noise labels, this indirect approach is prone to identify users' interested items as noise.
%The research optimizes the model by predicting the ground truth item, which indirectly guides the model to identify potential noise items for unsupervised denoising. However, the identified noise is not directly assigned true labels, and this indirect guidance may lead to errors in noise identification, reducing the denoising accuracy. 
% Additionally, labeling true noise requires significant human and material resources, making supervised denoising challenging to implement.
\item The research relies on the ground truth item to guide the model in identifying noise items within the sequence, focusing on item-level denoising. However, the ground truth item not only represents specific items the user is interested in but also reflects their broader interests. For example, when a user clicks on various sports-related items like ``football,'' ``sports shoes,'' and ``treadmills,'' these items indicate not just specific preferences but also a higher-level interest in ``sports.'' By considering this higher-level interest granularity, the model more effectively and broadly remove irrelevant noise, a factor often overlooked in previous methods.
\end{enumerate*}

In this paper, we propose a \textbf{M}ulti-\textbf{G}ranularity \textbf{S}equence \textbf{D}enoising with \textbf{W}eakly \textbf{S}upervised \textbf{S}ignal for Sequential Recommendation (MGSD-WSS).
It employs the Target-aware Sequence Encoding module to introduce weakly supervised noise signals and combines item-granularity and interest-granularity denoising modules to hierarchically remove noise.
\begin{enumerate*}[label=(\roman*)]
\item \textbf{Regarding the Target-aware Sequence Encoding module}, MGSD-WSS randomly introduces irrelevant items as noise into the historical interaction sequence to construct an augmented historical interaction sequence with noise annotations.
% We then train the model using a Multiple Gaussian-kernel Perceptron, which effectively identifies noise in the augmented historical interaction sequence through Gaussian kernel functions, generating denoised sequence representations.
We then train the model using a Multiple Gaussian-kernel Perceptron, which effectively enhances the sequence representation of historical interactions through Gaussian kernel functions and the target item.
\item \textbf{For the item-granularity denoising module}, it learns the denoised sequence representations and employs a noise-weighted contrastive learning approach to effectively avoid noise interference at the item granularity.
\item \textbf{For the interest-granularity denoising module}, it extracts the target interest representations from the ground truth item using a target-aware interest attention and constructs a user interest representation with noise weights from the augmented historical interaction sequence using an interest-wise weight generator. Based on these two representations, the module uses noise-weighted contrastive learning to effectively avoid noise interference at the interest granularity.
\end{enumerate*}

Extensive experiments on five datasets\cite{zhang2022hierarchical, zhang2024ssdrec} show that the proposed method significantly outperforms the state-of-the-art sequence recommendation and denoising models. Further analysis reveals that the advantages of our method stem from weakly supervised signals and the hierarchical denoising modules. Overall, our contributions are as follows:
\begin{itemize}  
\item We introduce labeled weakly supervised signals to directly train the model for noise identification, overcoming the inaccuracies of previous unsupervised methods. To the best of our knowledge, we are the first to apply weak supervision for denoising in sequential recommendation, offering a novel perspective to denoising techniques.

\item We use a hierarchical denoising module with noise-weighted contrastive learning to effectively remove noise at both the item and interest granularities, addressing the limitation of only removing item-level noise.

\item Extensive experiments and analyses demonstrate that our model outperforms state-of-the-art sequence recommendation and denoising models, showcasing its robust performance.

\end{itemize}

\begin{figure*}[htbp]  %当前位置排版，放不下可以在本页顶部或者底部
      \centering  %居中
      \includegraphics[width=1\textwidth]{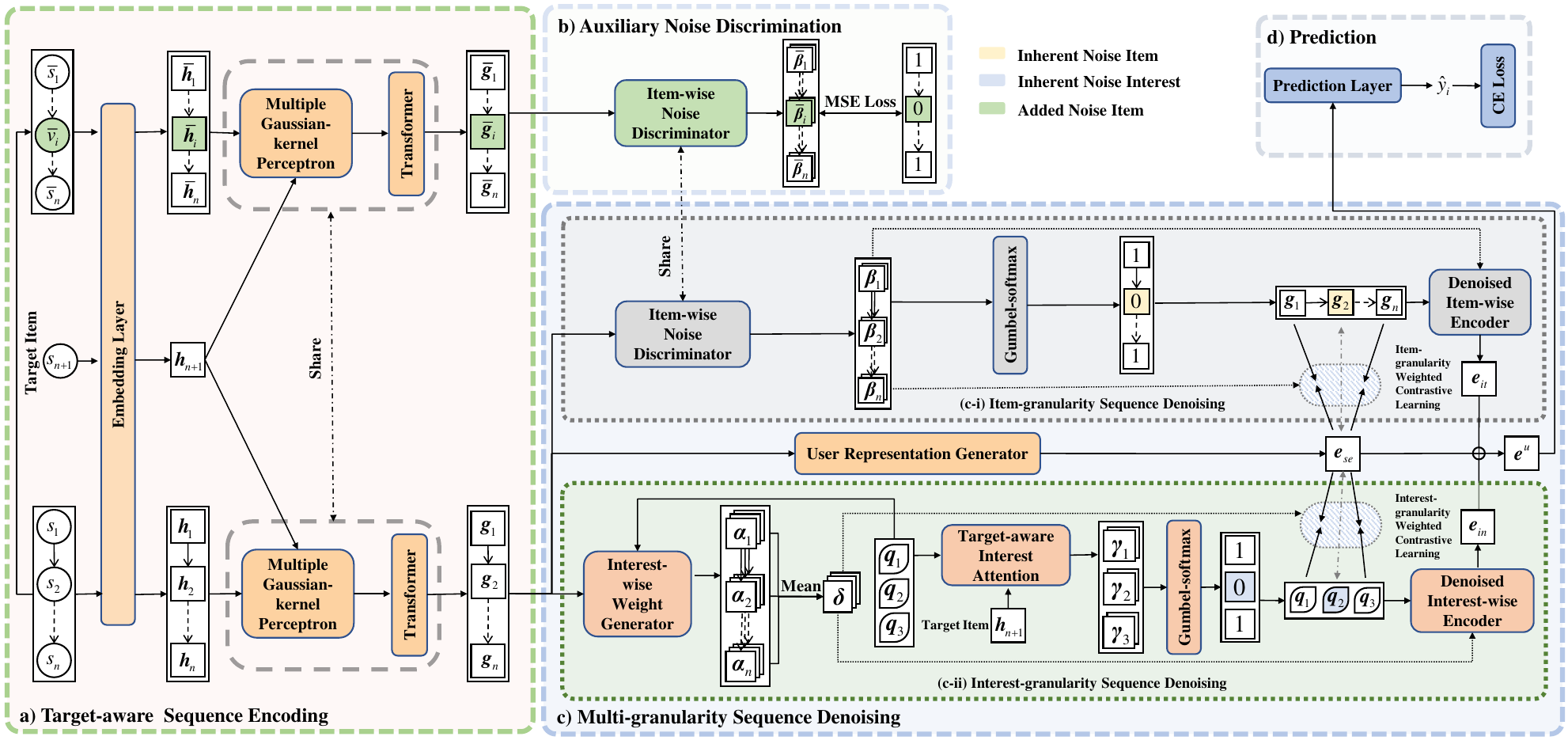} %photo是文件名，如果当前文件夹下没有同名不同格式的文件，比如photo.png就不需要后缀名啦
      \caption{The architecture of our proposed model MGSD-WSS.} 
      \label{fig:overall_framework} %指定给图片一个标签
\end{figure*}

\section{Related work}
\subsection{Sequential Recommendation}
The task of sequential recommendation involves modeling user preferences from historical interaction sequences to recommend items of potential interest. Early approaches in this field use Markov Chains (MCs) to model dynamic patterns in users' historical interactions, enabling personalized recommendations\cite{rendle2010factorizing, yang2020hybrid}. With the advancement of deep learning, various neural networks are integrated into sequence recommendation tasks to enhance item embeddings and extract user intent.
For example, Hidasi et al.\cite{hidasi2015session} propose a recurrent neural network (RNN) to capture temporal dependencies in historical interaction sequences, while Sak et al.\cite{sak2014long} introduce long short-term memory (LSTM) networks to address long-term dependencies in traditional RNN architectures. Tang et al.\cite{tang2018personalized} develop a convolutional neural network (CNN)-based approach to capture short-term user interests effectively. To prioritize critical information, Li et al.\cite{li2017neural} implement cognitive attention mechanisms inspired by human perception.
Graph neural networks (GNNs) are also widely adopted\cite{dong2022heterogeneous}. For instance, Wu et al.\cite{wu2019session} leverage graph structures to propagate and aggregate information across nodes and edges, capturing both local and global relational patterns. Recent studies further enhance model adaptability across domains and mitigate data sparsity by integrating external knowledge graphs\cite{liu2023knowledge} and cross-domain information\cite{xiao2023proxy}. Additionally, Zhou et al.\cite{zhou2024disentangled} employ multimodal learning frameworks to process heterogeneous data types and holistically model user behavior.
\subsection{Denoising Methods}
In the sequential recommendation task, historical interaction sequences inevitably contain noisy items\cite{tolomei2019you, zhang2020practical, chen2019data}. To address this, researchers have developed soft-denoising techniques that assign lower weights to unreliable interactions, reducing noise while preserving valuable behavioral signals.

Li et al.\cite{li2020time} propose a time-aware self-attention mechanism that considers the position and time intervals between items during the prediction process. Luo et al.\cite{luo2020collaborative} identify neighboring sessions and assign weights to collaborative items using self-attention networks. Yuan et al.\cite{yuan2021dual} introduce multiple self-attention networks to model the significance of each item in relation to the target item and apply an adaptive sparse transformation function to minimize the impact of noise items. Zhou et al.\cite{zhou2023attention} improve item attention weights by using spatial and adversarial calibrators, adjusting the weights based on spatial information and item contributions. Zhu et al.\cite{zhu2025preference} propose a preference-driven denoised fusion module that integrates denoised frequency and variational representations, aligning fused embeddings with true user preferences while minimizing residual noise.

However, soft-denoising methods have limitations in fully eliminating noise, leading to the exploration of hard-denoising techniques. These methods aim to remove noise completely by generating binary noise detection signals. For example, Sun et al.\cite{sun2021does} model the loss and uncertainty associated with unreliable items using a Gaussian distribution. Tong et al.\cite{tong2021pattern} use reinforcement learning to determine the relevance of each item in the sequence, framing the denoising task as a Markov decision process. Chen et al.\cite{chen2022denoising} introduce a denoising strategy that removes attention values from noise items via a trainable binary mask in each self-attention layer. Zhang et al.\cite{zhang2022hierarchical} present a hierarchical sequence denoising approach that identifies inconsistent items using user-level intent signals and sequence-level context signals. Zhang et al.\cite{zhang2024ssdrec} apply sequence augmentation followed by a hierarchical hard-denoising module that uses Gumbel-Softmax selection to identify and remove noisy interactions, preventing over- and under-denoising.

These methods guide the model to remove noise using the ground truth item rather than true noise labels, achieving unsupervised denoising at the item granularity. They struggle to accurately denoise without true noise labels and fail to effectively denoise at the broader interest granularity. In contrast, we introduce weakly supervised signals and interest-granularity denoising methods, which more effectively remove noise from historical interaction sequences.

\subsection{Contrastive Learning}
The fundamental idea behind contrastive learning is to cluster samples from the same category in the embedding space while dispersing those from distinct categories. This approach has been proven effective in extracting high-quality representations from intricate and multifaceted features, thereby elevating model performance. For instance, Qiu et al.\cite{qiu2022contrastive} employ techniques like dropout to generate positive samples from original user behavior sequences, addressing representation degeneration in sequential recommendation systems. Chen et al.\cite{chen2022intent} introduce intent contrastive learning, aligning user intents with augmented sequences to bolster sequential recommendations. Duan et al.\cite{duan2023clsprec} leverage contrastive learning to differentiate user's long-term and short-term preferences, aiming to enhance the precision of next-item recommendations. Zhu et al.\cite{zhu2024high} utilize high-level preferences as positive exemplars in contrastive learning to refine multi-interest sequential recommendation. Nonetheless, these contrastive learning methods treat all samples equally, ignoring the different importance of different samples to the model. Therefore, we propose a weighted contrastive learning framework to rectify this limitation.

\subsection{Data Argument for Recommendation}
In sequential recommendation, data augmentation is pivotal in bolstering model generalization and robustness. Techniques like random masking, reordering, or inserting noise into sequences are commonly employed for this purpose\cite{dang2024repeated, wei2024llmrec, zhou2024contrastive, wu2021self}. These methods aim to generate diverse yet semantically similar sequences for contrastive learning.
Dang et al.\cite{dang2024repeated} propose repeated padding as a data augmentation strategy to boost model performance by enriching input data for sequential recommendation.
Wei et al.\cite{wei2024llmrec} tackle data sparsity in recommendation systems by introducing graph augmentation with large language models.
Wang et al.\cite{wang2022explanation} leverage explanation-guided contrastive learning for sequential recommendation. This method derives positive and negative sequences based on item importance.
Wu et al.\cite{wu2021self} explore self-supervised learning on user-item graphs to improve the accuracy and robustness of graph convolutional network recommendations.
% Our model enhances noise recognition capability and improves user representation quality by adding noise items into the user's historical interactions to generate the noise supervised signals.
The lack of noise supervised signals has undermined the ability of existing methods to accurately distinguish between noise items and non-noise items (i.e., those that reflect the user's true intentions).
To address this, we introduce an auxiliary task that generates a noise supervised signal for identifying noise items.

\section{Preliminaries} 
\subsection{Problem Formulation} 
Suppose we have a set of users $\mathcal{U} \in \{u_{1}, u_{2}, \cdots, u_{\vert\mathcal{U}\vert}\}$ and items $\mathcal{V} \in \{v_{1}, v_{2}, \cdots, v_{\vert\mathcal{V}\vert}\}$, respectively.
Each user $u \in \mathcal{U}$ is associated with a historical interaction sequence $S=[s_{1}, s_{2}, \cdots, s_{n}]$ in chronological order, where $\textit{n}$ indicates the length of the sequence, and $s_{i}$ denotes the $\textit{i}$-th item interacted by $u$.
The purpose of sequential recommendation is to utilize the sequence of interactions $S$ to predict the item that the user $u$ is most likely to interact with at the ($n$+1)-th step, denoted as $p(s_{n+1}|s_{1:n})$. 

%\zhu{20241114,14:53}

\subsection{Sequence Data Augmentation}
In real-world scenarios, the noise within historical interaction sequences leads to suboptimal model results.
The absence of noise supervised signals poses a significant challenge in effectively removing noise items from these historical interaction sequences.
To address this, we introduce an augmentation strategy to bridge the gap caused by the absence of noise supervised signals. Specifically, we randomly select $t$ different items $\{\bar{v}_{i}\}_{i=1}^{t}$ from the item set $ \mathcal{V}$ to represent noises.
These noise items are inserted at arbitrary positions into the original historical interaction sequence $S=[s_{1}, s_{2}, \cdots, s_{n}]$, thereby creating an augmented sequence $\bar{S}=[\allowbreak\bar{s}_{1}, \allowbreak\bar{s}_{2}, \allowbreak\cdots, \allowbreak\bar{v}_{i}, \allowbreak\cdots, \allowbreak\bar{s}_{n+t}]$. Then, we obtain supervised signals $\bar{\boldsymbol{z}}=[\allowbreak\bar{z}_{1}, \allowbreak\bar{z}_{2}, \allowbreak\cdots, \allowbreak\bar{z}_{i}, \allowbreak\cdots, \allowbreak\bar{z}_{n+t}]$ to indicate the location of noise within the sequence. 
Note that we truncate the preceding items of the augmented sequence  and keep $\bar{S}$ having the  same length as $S$.

\section{Methodology}  
In this section, we provide a detailed explanation of our proposed model, MGSD-WSS, as illustrated in Figure \ref{fig:overall_framework}. The model consists of three key components: Target-aware Sequence Encoding, Auxiliary Noise Discrimination and Multi-granularity Sequence Denoising.
%the multiple gaussian-kernel perceptron enhances the robustness of the model; the multi-granularity sequence denoising filters out noise at both interest and item granularities. Additionally, the auxiliary noise discrimination employs explicit noise guidance signals to improve the model's ability to identify noise.

\subsection{Target-aware Sequence Encoding}
In this module, we attempt to learn informative representations for items in both the original sequence and the augmented sequence.  To the end, we first obtain the embedding of each item in the two sequences by employing an embedding layer. Then, we attempt to map the two item sequences into a common feature space by leveraging a  shared Multiple Gaussian-kernel Perceptron (MGP) module together with a transformer. The MGP module is developed to capture different levels of matching features\cite{wu2023adversarial}, and the transformer is introduced to capture long-range contextual information of items.  
% transformers excel at capturing long-range dependencies and establishing meaningful relationships between words, enabling them to generate high-quality, context-aware text

\subsubsection{Embedding Layer}
We feed the user's historical interaction sequence $S$ into the embedding layer, and obtain its corresponding item embedding matrix $\boldsymbol{H} \in \mathbb{R}^{n \times d}$, where $n$ represents the sequence length and $d$ denotes the embedding size. Analogously, for the augmented user's historical interaction sequence $\bar{S}$, we employ $\boldsymbol{\bar{H}} \in \mathbb{R}^{n \times d}$ as the corresponding item embedding matrix. To capture the positional information of different items in $S$, we utilize a trainable parameter matrix  $\boldsymbol{P} \in \mathbb{R}^{n \times d}$ to  signify the specific position of each item $\{s_{i}\}_{i=1}^{n}$ within $S$. For simplicity, $\boldsymbol{h}_{i} \in \boldsymbol{H}$ and $\boldsymbol{\bar{h}}_{i} \in \boldsymbol{\bar{H}}$ represent the item embedding representations of $s_{i} \in S$ and $\bar{s}_{i} \in \bar{S}$, respectively, whereas $\boldsymbol{p}_{i} \in \boldsymbol{P}$ and $\boldsymbol{\bar{p}}_{i} \in \boldsymbol{\bar{P}}$ denote the corresponding positional embedding.

%\zhu{(20241127: Need to add definition of $\bar{P}$)}

\subsubsection{Multiple Gaussian-kernel Perceptron (MGP)}

% 动机：使用不同的高斯核来计算各个表示的重要性
%      提取到不同域的信息
%\zhu{(20241204: Gaussian-kernal function)}
%为了更好的利用噪音增强序列\bar{S}来引导模型识别原序列S中的噪音items，我们设计了一种共享参数的Multiple Gaussian-kernel Percentron (MGP)将原序列S和噪音增强序列\bar{S}映射到公共表示空间。 
%To  transfer the embedding   ommon representation space
To better leverage the augmented sequence $\bar{S}$ to assist the model in identifying the noisy items in the original sequence $S$, we propose a Multiple Gaussian-kernel Perceptron (MGP) to map both sequences into a common representation space with shared-parameters. 
% Specifically, we first compute the relevance score vector for items in a sequence, and then utilize multiple Gaussian kernels to model different matching level of relevance by transforming it into kernel features. 
% Gaussian kernel has been proved to be effective in capturing matching features\cite{jin2022towards,wu2024adversarial}, and we employ 
%
As the same MGP operation are applied to both sequences, we concentrate on processing the augmented sequence $\bar{S}$. Specifically, we first utilize the 
cosine similarity to compute the correlation between the target item $\boldsymbol{\bar{h}}_{n+1}$ and each item $\boldsymbol{\bar{h}}_{i}$ in $\bar{S}$, resulting in a target-aware relevance score $\bar{r}_{i} \in \mathbb{R}$. Notably, in this study, $\boldsymbol{\bar{h}}_{n+1}$ is used during the training phase, whereas $\boldsymbol{\bar{h}}_{n}$ is employed during validation and testing phases.It can be denoted as follows: 
\begin{align}
    \bar{r}_{i}=\cos\left(\boldsymbol{\bar{h}}_{n+1},\ \boldsymbol{\bar{h}}_{i}\right).
\end{align}

We employ $k$ Gaussian kernels to transfer the target-aware relevance score into a refined score $\boldsymbol{\tilde{r}}_{i} \in \mathbb{R}^{k}$, which is then applied to weight and aggregate the items $\boldsymbol{\bar{h}}_{i}$ in the sequence. The precess is defined  as follows:
\begin{align}
    \boldsymbol{\tilde{r}}_{i} &=\left[\tilde{r}_{i1},\ \tilde{r}_{i2},\ \cdots,\ \tilde{r}_{ik}\right],\\
    \tilde{r}_{ij} &=\exp\left(-\frac{\left(\bar{r}_{i}-\mu_{j}\right)^{2}}{2\sigma_{j}^{2}}\right),\\
    \boldsymbol{\hat{h}}_{i} &=\sum_{j=1}^{k}\tilde{r}_{ij}\boldsymbol{\bar{h}}_{i},
\end{align}
where $\mu_{j}$ and $\sigma_{j}$  representing the mean and variance of the $j$-th kernel, respectively. These kernels are adept at capturing information from various similarity regions of the items\cite{xiong2017end}. 

After that, a Transformer\cite{zhou2020s3, xu2023group} is utilized to model the enriched representations of the items $\boldsymbol{\hat{h}}_{i} \in \boldsymbol{\hat{H}}$, incorporating richer contextual information. 
\begin{align}  
    \boldsymbol{G}&=\operatorname{Transformer} \left(\boldsymbol{\hat{H}}+\boldsymbol{P}\right),
\end{align}
where $\boldsymbol{\bar{G}}=\left[\boldsymbol{\bar{g}}_{1},\ \boldsymbol{\bar{g}}_{2},\ \cdots,\ \boldsymbol{\bar{g}}_{n}\right] \in \mathbb{R}^{n \times d}$ designates the hidden representation matrix derived from the final layer of the Transformer encoder, $\boldsymbol{\hat{H}}$ and $\boldsymbol{P}$ respectively represent the item embedding matrix and its corresponding position embedding matrix. 
Analogously, for the original historical interaction sequence $S$, a same process yields an enriched item hidden representation matrix denoted as  
$\boldsymbol{G}=\left[\boldsymbol{g}_{1},\ \boldsymbol{g}_{2},\ \cdots,\ \boldsymbol{g}_{n}\right] \in \mathbb{R}^{n \times d}$. 

\subsection{Auxiliary Noise Discrimination}
The lack of noise supervised signals has undermined the ability of existing methods to accurately distinguish between noise items and non-noise items (i.e., those that reflect the user's true intentions), resulting in suboptimal performance.
To address this, we introduce the Auxiliary Noise Discrimination module. Specifically, a shared item-wise noise discriminator module is employed to detect noise items within the augmented historical interaction sequence $\boldsymbol{\bar{G}}=\left[\boldsymbol{\bar{g}}_{1},\ \boldsymbol{\bar{g}}_{2},\ \cdots,\ \boldsymbol{\bar{g}}_{n}\right] \in \mathbb{R}^{n \times d}$. Additionally, a shard target-aware gaussian kernel sequence encoding module is utilized to enrich items' multi-granular information. To minimize the disparity between the noise item detection signals $\boldsymbol{\bar{\beta}}_{i} = [\bar{\beta}_{i}^{0},\ \bar{\beta}_{i}^{1}]$ and the noise supervised signals $\bar{z}_{i} \in \left[0,\ 1\right]$, we apply the Mean Squared Error (MSE) loss:
\begin{align}
    &\boldsymbol{\bar{\beta}}_{i} = \operatorname{Softmax} \left(\left(\operatorname{ReLU}\left(\boldsymbol{\bar{g}}_{i}\boldsymbol{W}_{1} + \boldsymbol{b}_{1}\right)\right) \boldsymbol{W}_{2}\right), \\
    &\mathcal{L}_{MSE} = \frac{1}{n} \sum_{i=1}^{n}\left(\bar{\beta}_{i}^{0}-\bar{z}_{i}\right)^{2}, \label{eq: MSE}
\end{align}
where $\boldsymbol{W}_{1} \in \mathbb{R}^{d \times d}$, $\boldsymbol{W}_{2} \in \mathbb{R}^{d \times 2}$, and $\boldsymbol{b}_{1} \in \mathbb{R}^{d}$ are learnable parameters. %$n$ represents the sequence length of the augmented historical interaction sequence $\bar{S}$.

\subsection{Multi-granularity Sequence Denoising}
\subsubsection{User Representation Generator}
To obtain the sequence-based user representation $\boldsymbol{e}_{se} \in \mathbb{R}^{d}$, we utilize a sequence encoder to model the item representation matrix $\boldsymbol{G}$. This can be expressed as follows:
\begin{align}
    \boldsymbol{e}_{se}=\text{F}\left(\boldsymbol{G}\right),
\end{align}
where $\text{F}\left(\cdot\right)$ can be any prevalent sequence recommendation model, such as BERT4Rec\cite{sun2019bert4rec} or GRU4Rec\cite{hidasi2015session}.

\subsubsection{Item-granularity Sequence Denoising}
Considering that the user sequences inherently contain noise, such as accidental clicks, it is suboptimal to model these sequences directly without any denoising process. To address this, we utilize the shared item-wise noise discriminator to identify noise items within the user's historical interaction sequence. Finally, we use the non-noise items to construct item-granularity user representations $\boldsymbol{e}_{it}$.

% Considering that the user sequences inherently contain noise, such as accidental clicks, it is suboptimal to model these sequences directly without any denoising process. To address this, we utilize a multi-layer perceptron (MLP) to identify noise items within the user's historical interaction sequence. We then use the non-noise items to construct item-level user representations $\boldsymbol{e}_{it}$.

\textbf{Item-wise Noise Discriminator.}
To efficiently distinguish the noise items in the sequences, we utilize the noise detection signal $\boldsymbol{\beta}_{i} \in \mathbb{R}^{2}$, generated by the shared item-wise noise discriminator, to discern whether $\boldsymbol{g}_{i}$ corresponds to a noisy item. The specific procedure is as follows:
\begin{align}
    \boldsymbol{\beta}_{i}=\operatorname{Softmax} \left(\left(\operatorname{ReLU}\left(\boldsymbol{g}_{i}\boldsymbol{W}_{1} + \boldsymbol{b}_{1}\right)\right) \boldsymbol{W}_{2}\right),
\end{align}
where $\boldsymbol{W}_{1} \in \mathbb{R}^{d \times d}$, $\boldsymbol{W}_{2} \in \mathbb{R}^{d \times 2}$, and $\boldsymbol{b}_{1} \in \mathbb{R}^{d}$ are learnable parameters.
Subsequently, the Gumbel-softmax function is employed to transform $\boldsymbol{\beta}_{i} = \left[\beta_{i}^{0},\ \beta_{i}^{1}\right] \in \left[0,\ 1\right]$ into a binary hard value $\boldsymbol{\tilde{\beta}}_{i} = \left[\tilde{\beta}_{i}^{0},\ \tilde{\beta}_{i}^{1}\right] \in \{0,\ 1\}$, thereby assessing the reliability of item $\boldsymbol{g}_{i}$. It is noteworthy that only the first dimension of the hard signal $\boldsymbol{\tilde{\beta}}_{i}$ is utilized. Specifically, $\tilde{\beta}_{i}^{0}=1$ signifies that $\boldsymbol{g}_{i}$ represents a non-noise item, whereas $\tilde{\beta}_{i}^{0}=0$ indicates a noise item. The detailed procedures are as follows:
\begin{align}
    \tilde{\boldsymbol{\beta}}_{i} &= \operatorname{Gumbel-softmax}\left(\boldsymbol{\beta}_{i}, \tau\right), \\
    \tilde{\beta}_{i}^{0} &= \frac{\exp\left(\log\left(\beta_{i}^{0}\right) + g_{0}\right)/\tau}{\sum_{j=0}^{1}\exp\left(\log\left(\beta_{i}^{j}\right) + g_{j}\right) / \tau},
\end{align} 
where $g_{j}$ represents a disturbance sampled from the Gumbel distribution, aiming to bolster the model's robustness. The temperature parameter is defined as $\tau > 0$. As $\tau \rightarrow 0$, $\Tilde{\boldsymbol{\beta}}_{i}$ is approximated by a one-hot vector. As $\tau \rightarrow \infty$, $\Tilde{\boldsymbol{\beta}}_{i}$ is approximated by a uniform distribution. As $\tau \rightarrow 1$, the Gumbel-softmax function gradually converges towards the standard Softmax function\cite{zhang2022hierarchical}.

\textbf{Denoised Item-wise Encoder.}
The binary hard signal  $\boldsymbol{\tilde{\beta}}^{0}=[\tilde{\beta}^{0}_{1},\ \tilde{\beta}^{0}_{2},\ \cdots,\ \tilde{\beta}^{0}_{n}] \in \mathbb{R}^{n}$ is employed to filter out noise items from the user's original historical interaction sequence, forming a denoised  hidden representation matrix $\boldsymbol{C}=\left[\boldsymbol{c}_{1}, \boldsymbol{c}_{2}, \cdots, \boldsymbol{c}_{t}\right] \in \mathbb{R}^{t \times d}, t \leq n $. Subsequently, we utilize a Transformer encoder to model $\boldsymbol{C}$ and obtain the item-wise user representation $\boldsymbol{e}_{it}$: 
\begin{align}
    \boldsymbol{C} &= \operatorname{Reorganize}\left(\left(\boldsymbol{1} - \boldsymbol{\tilde{\beta}}^{0}\right) \cdot \boldsymbol{G}\right), \\
    \boldsymbol{e}_{it} &= \operatorname{Transformer}\left(\boldsymbol{C}\right),
\end{align}
where $\operatorname{Reorganize}\left(\cdot\right)$ is responsible for reorganizing the sequence of user interactions by discarding noise items.

\textbf{Item-granularity Weighted Contrastive Learning.}
%使用gumbel-softmax得到的噪音信号 可以看成是弱监督信号。
After gathering the positive and negative samples via the items noise detection signal $\{\tilde{\beta}_{i}^{0}\}_{i=1}^{n}$, we employ the contrastive learning to align samples. As described in Figure \ref{fig:overall_framework}(c-i), $\boldsymbol{g}_{1}$ and $\boldsymbol{g}_{2}$ respectively represent non-noise and noise items, we aim to maximize the consistency between positive samples $\left(\boldsymbol{e}_{se}, \ \boldsymbol{g}_{1}\right)$ and minimize the consistency between negative samples $\left(\boldsymbol{e}_{se}, \ \boldsymbol{g}_{2}\right)$.

As traditional contrastive learning methods usually treat each sample with equal weights, they neglect the varying noise levels within different samples. To incorporate the noise levels as the weights of their corresponding samples, and introduce a weighted framework of contrastive learning. To be specific, the  positive and negative samples are respectively weighted by $\beta_{i}^{0}$ and $1-\beta_{i}^{0}$:
\begin{flalign}    
    &\omega\left(\boldsymbol{g}_{i}\right) = \left\{
	\begin{aligned}
		\beta_{i}^{0} &  & \boldsymbol{g}_{i} \in \boldsymbol{G}^{+},\\
		1 - \beta_{i}^{0} &  & \boldsymbol{g}_{i} \in \boldsymbol{G}^{-},
	\end{aligned}
	\right.&
    % \end{split}
\end{flalign}
where $\boldsymbol{G}^{+}$ and $\boldsymbol{G}^{-}$ denote the representations of non-noise items and noise items within the sequence, respectively.
The loss of the item-granularity weighted contrastive learning is defined as follows:
\begin{flalign}
    % \begin{split} \label{eq: ITSCL}
    &\mathcal{L}_{ITSCL} \!=\! -\frac{1}{\!|\boldsymbol{G}^{+}\!|} 
    \!\sum_{\boldsymbol{g}_{i} \in \boldsymbol{G}^{+}}\!\log \frac{\omega\!\left(\boldsymbol{g}_{i}\!\right) \!\cdot \exp\!\left(\text{sim}\!\left(\boldsymbol{e}_{se}, \boldsymbol{g}_{i}\!\right) \!/ \tau\!\right)}{\!\sum\limits_{\boldsymbol{g}_{j} \in \boldsymbol{G}} \! \omega\!\left(\boldsymbol{g}_{j}\!\right) \!\cdot \exp\!\left(\text{sim}\!\left(\boldsymbol{e}_{se}, \boldsymbol{g}_{j}\!\right) / \tau\!\right)},&  \label{eq: ITSCL}  
\end{flalign}
where $|\boldsymbol{G}^{+}|$ represents the number of non-noise items in the sequence.

\subsubsection{Interest-granularity Sequence Denoising}
Considering that user interests change dynamically in real-world scenarios, traditional item-based sequence denoising methods\cite{li2020time, tong2021pattern, zhou2023attention} 
focus on item-level denoising, i.e., removing individual noisy items within an interaction sequence. However, the interacted items sequence of a user is mainly determined by her inherent interest preference.
% focus solely on the noise identification of specific items, \li{neglecting the filtering of noisy or irrelevant interests within historical interaction sequences.} 
Therefore, how to explore and denoise these interests is crucial to improve the performance of the recommender system.
However, due to the diversity of user interests, traditional methods that only model  users' short- and long-term interests are becoming unreliable. To resolve this problem, users' multiple interests are modeled by considering the distinct features of the items in the sequence and utilizing target-aware attention mechanisms to filter out noisy interests. Finally, the filtered high-quality user interests are used to model the interest-granularity user representation $\boldsymbol{e}_{in} \in \mathbb{R}^{d}$, in order to obtain the topics or categories that truly reflect the users' interests.

Specifically, most existing denoising approaches 
focus on item-level denoising, i.e., removing individual noisy items within an interaction sequence. However, the interacted items sequence of a user is mainly determined by her inherent interest preference, thus a broader interest-level denoising can better reflect user information preference. As the two denoising strategies belong to different granularities, performing denoising across them can comprehensively learn more robust and accurate user preference.

\textbf{Interest-wise Weight Generator.}
To capture a user's multi-interest preferences, we introduce a learnable interest representation matrix $\boldsymbol{Q}=\left[\boldsymbol{q}_{1},\ \boldsymbol{q}_{2},\ \cdots,\ \boldsymbol{q}_{m}\right] \in \mathbb{R}^{m \times d}$, where $m$ represents the number of the user interests. We then compute the relevance score $\boldsymbol{\alpha}_{i} \in \mathbb{R}^{m}$ between each item $\boldsymbol{g}_{i}$ in the sequence and $\boldsymbol{Q}$:
\begin{align}
    \boldsymbol{\alpha}_{i} &=\operatorname{Softmax}\left(\boldsymbol{Q}\left(\operatorname{ReLU}\left(\boldsymbol{g}_{i}\boldsymbol{W}_{3} + \boldsymbol{b}_{2}\right)\right) \right), 
\end{align}
where   $\boldsymbol{W}_{3} \in \mathbb{R}^{d\times d}$ and $\boldsymbol{b}_{2} \in \mathbb{R}^{d}$ are learnable parameters. 
Subsequently, we obtain interest distribution $\boldsymbol{\delta}=\left[\delta_{1},\ \delta_{2},\ \cdots,\ \delta_{m}\right] \in \mathbb{R}^{m}$ of the current user across $m$ interests as follows: 
\begin{align} 
    \boldsymbol{\delta} &= \frac{1}{n} \sum_{i=1}^{n}\boldsymbol{\alpha}_{i}.
\end{align}

\textbf{Target-aware Interest Attention.}
After we capture the user multiple interests, it is uncertain whether each interest contributes equally to improving the model's recommendation performance. Therefore, we introduce the target item information to assess the reliability distribution $\boldsymbol{\gamma}_{l} = [\gamma_{l}^{0},\ \gamma_{l}^{1}]$ of the \textit{l}-th user interest $\boldsymbol{q}_{l}$ as follows: 
\begin{align}
    \boldsymbol{\hat{q}}_{l}&=\boldsymbol{q}_{l}\boldsymbol{W}_{4},\\
    \boldsymbol{\gamma}_{l} &= \operatorname{Softmax}\left(\left[\boldsymbol{\hat{q}}_{l} \| \boldsymbol{h}_{n+1} \|
    \boldsymbol{\hat{q}}_{l} - \boldsymbol{h}_{n+1} \|
    \boldsymbol{\hat{q}}_{l} \odot \boldsymbol{h}_{n+1}\right] \boldsymbol{W}_{5} \right),
\end{align}
where $\boldsymbol{W}_{4} \in \mathbb{R}^{d \times d}$ and $\boldsymbol{W}_{5} \in \mathbb{R}^{4d \times 2}$ are learnable parameters,  $\odot$ represents element-wise product. $\gamma_{l}^{0}$ and $\gamma_{l}^{1}$ indicate the probabilities of the reliability and unreliability of the \textit{l}-th user interest, respectively. 
Then, we employ the Gumbel-softmax function to transform $\boldsymbol{\gamma}_{l}$ into a binary hard value $\boldsymbol{\tilde{\gamma}}_{l} = [\tilde{\gamma}_{l}^{0},\ \tilde{\gamma}_{l}^{1}]$:
\begin{align}
    \boldsymbol{\tilde{\gamma}}_{l}=\operatorname{Gumbel-softmax} \left(\boldsymbol{\gamma}_{l}, \ \tau\right).
\end{align}

Note that we solely rely on the first dimension of the hard signal $\boldsymbol{\tilde{\gamma}}_{l}$ for our determination. Specifically, $\tilde{\gamma}_{l}^{0} = 1$ denotes $\boldsymbol{q}_{l}$ is a non-noise interest, otherwise $\tilde{\gamma}_{l}^{0} = 0$ indicates $\boldsymbol{q}_{l}$ is a noise interest.

\textbf{Denoised Interest-wise Encoder.}
To derive the interest-wise user representation $\boldsymbol{e}_{in}$, we leverage the $l$-th interest representation $\boldsymbol{q}_{l} \in \boldsymbol{Q}$, 
the corresponding user interest weight $\delta_l \in \boldsymbol{\delta}$ and the generated noise signal $\tilde{\gamma}_{l}^{0}$ to eliminate noise interests. Formally, we have:
\begin{align}
    \boldsymbol{e}_{in} =\sum^{m}_{l=1}\delta_l \tilde{\gamma}_{l}^{0} \boldsymbol{q}_{l}.
\end{align}

\textbf{Interest-granularity Weighted Contrastive Learning.}
Analogous to the aforementioned item-granularity weighted contrastive learning, we utilize the user interests noise  signal $\{\tilde{\gamma}_{l}^{0}\}_{l=1}^{m}$ to identify positive and negative samples. Subsequently, we apply user interest weights $ \{\delta_l\}_{l=1}^{m}$ to weight these samples. As illustrated in Figure \ref{fig:overall_framework}(c-ii),  $\boldsymbol{q}_{2}$ signifies a noisy interest, whereas $\boldsymbol{q}_{1}$ is a relevant interest. The contrastive learning objective is to maximize the consistency between the positive sample $(\boldsymbol{e}_{se}, \ \boldsymbol{q}_{1})$ and minimize the consistency between the negative sample $(\boldsymbol{e}_{se}, \ \boldsymbol{q}_{2})$:
\begin{flalign} \label{eq: INSCL}
    % \begin{split}
    &\mathcal{L}_{INSCL} = -\frac{1}{|\boldsymbol{Q}^{+}|} 
    \sum_{\boldsymbol{q}_{l} \in \boldsymbol{Q}^{+}}\log \frac{\delta_l \cdot \exp\left(\text{sim}\left(\boldsymbol{e}_{se}, \boldsymbol{q}_{l}\right) / \tau\right)}{\sum\limits_{\boldsymbol{q}_{j} \in \boldsymbol{Q}}\delta_{j} \cdot \exp\left(\text{sim}\left(\boldsymbol{e}_{se}, \boldsymbol{q}_{j}\right) / \tau\right)},&
    % \end{split}
\end{flalign}
where $\text{sim}\left(\cdot \right)$ represents the cosine similarity function,  $\tau$ and $|\boldsymbol{Q}^{+}|$ denote a temperature parameter and the number of positive user interests, respectively.

\subsection{Prediction and Model Optimization}
In this module, we merge the interest-wise user representation $\boldsymbol{e}_{in}$, item-wise user representation $\boldsymbol{e}_{it}$, and sequence-wise user representation $\boldsymbol{e}_{se}$ to form the final user representation $\boldsymbol{e}^{u}$. We then compute the score between $\boldsymbol{e}^{u}$ and each item $v_{i} \in \mathcal{V}$ using dot product, which is utilized to  predict the probability distribution $\hat{y}_{i}$ for the next item. We employ the cross-entropy between the prediction and ground-truth as the loss function: 
\begin{align}
    &\boldsymbol{e}^{u} = \boldsymbol{e}_{in} + \boldsymbol{e}_{it} + \boldsymbol{e}_{se}, \\
    &\hat{y}_{i} = \frac{\exp\left(\boldsymbol{e}^{u}\boldsymbol{h}_{v_{i}}\right)} {\sum_{v_{j} \in \mathcal{V}}\exp\left(\boldsymbol{e}^{u}\boldsymbol{h}_{v_{j}}\right)},\\
    &\mathcal{L}_{REC} = -\sum_{i=1}^{|\mathcal{V}|}
    \left(
    y_{i}\log\left(\hat{y}_{i}\right) + 
    \left(1-y_{i}\right)\log\left(1-\hat{y}_{i}\right)
    \right),
\end{align}
where $y_{i}$ is the ground-truth for the $i$-th item.

Finally, the overall optimization loss for joint training of the recommendation task is defined as a combination of   all above losses:
\begin{align}
    \begin{split}
    \mathcal{L} = \frac{1}{|\mathcal{U}|}\sum\limits_{u=1}^{|\mathcal{U}|}
    &\left(
    \mathcal{L}_{REC} \right. 
    + \lambda_1\mathcal{L}_{ITSCL} 
    + \lambda_2\mathcal{L}_{INSCL} \\
    & \left. + \lambda_3\mathcal{L}_{MSE} \right) 
    + \rho\Vert\Theta\Vert_{2}^{2},
    \end{split}
\end{align}
where  $\lambda_1$, $\lambda_2$, $\lambda_3$ are hyperparameters to control the strengths of different losses, $\rho$ is a hyperparameter to control $L_2$ regularization, and $\Theta$ indicates the learnable model parameters.

% model end
%%%%========================================================%%%%
\section{Experiments}
% In this section, we compare  our proposed model with  various mainstream sequential recommendation models and state-of-the-art sequence denoising models on five real-world datasets. 

% Our experiments are aimed at answering the following research questions:
% \begin{itemize}
%     \item \textbf{RQ1:} When integrated with various mainstream sequential recommendation models, how does our approach perform in comparison to the original models? Furthermore, how does it compare to state-of-the-art sequence denoising models?
%     \item \textbf{RQ2:} What role do the distinct components of our method play in enhancing model performance?
%     \item \textbf{RQ3:} Within the framework of our approach, how effective is weighted contrastive learning compared to traditional contrastive learning?
% \end{itemize}

\begin{table}[]  
\caption{Statistics of the datasets.}
\scalebox{0.8}{
    \begin{tabular}{lccccc}
    \toprule
    Dataset   & \#Sequence    & \#Users   & \#Items   & \#Avg.length  & \#Sparsity  \\ \midrule
    ML-100k   & 99,287        & 944       & 1,350     & 105.29        & 92.21\%     \\ 
    Beauty    & 198,502       & 22,364    & 12,102    & 8.88          & 99.93\%     \\ 
    Sports    & 296,337       & 35,599    & 18,358    & 8.32          & 99.95\%     \\ 
    Yelp      & 316,354       & 30,432    & 20,034    & 10.40         & 99.95\%     \\ 
    ML-1M     & 999,611       & 6,041     & 3,417     & 165.50        & 95.16\%     \\
    \bottomrule
    \end{tabular}\label{tab:dataset}
}
\end{table}
\subsection{Dataset}
We conduct extensive experiments on five public benchmark datasets to evaluate the performance of our model. The details of these datasets are summarized in Table \ref{tab:dataset}.
\begin{itemize}
    \item \textbf{MovieLens}\footnote{https://movielens.org/}: This dataset comprises user ratings and reviews of movies, and is widely utilized in various sequential recommendation models. For our experiments, we utilize ML-100k and ML-1M.
    \item \textbf{Amazon-Beauty and Sports}\footnote{http://jmcauley.ucsd.edu/data/amazon}: The Amazon dataset contains user purchasing patterns across diverse products. We specifically choose  two subsets, i.e., Beauty and Sports, for our analysis.
    \item \textbf{Yelp}\footnote{https://www.yelp.com/dataset}: A widely used dataset in business recommendation research, which includes user reviews of numerous restaurants and bars. We utilize transaction records after January 1st, 2019.
\end{itemize}

For all datasets, user interactions are grouped and organized into chronological sequences. Following the methodologies  in\cite{zhang2022hierarchical, zhu2024multilevel}, we exclude inactive items and users which have fewer than five interactions. We employ the leave-one-out evaluation strategy and set a maximum sequence length of 200 for ML-1M and 50 for the remaining datasets to maintain experimental consistency.

\subsection{Evaluation Metrics}
We employ three commonly used evaluation metrics\cite{zhang2022hierarchical, zhu2024multilevel} to assess the performance of all methods, including \textit{Hit Ratio} (HR@$\{5,10,20\}$), \textit{Normalized Discounted Cumulative Gain} (NDCG@$\{5,10,20\}$), and \textit{Mean Reciprocal Rank} (MRR@20). 

\subsection{Baseline Models}
To validate the effectiveness of our proposed sequence denoising framework, we integrate MGSD-WSS with six base sequential recommendation models and conduct a comprehensive comparison with their original versions.
\begin{itemize}
    \item \textbf{GRU4Rec}\cite{hidasi2015session} introduces a tailored ranking loss function and gated recurrent unit (GRU) for session-based recommendation. 
    \item \textbf{NARM}\cite{li2017neural} utilizes a hybrid encoder with an attention mechanism to model the main purpose of the user. 
    \item \textbf{STAMP}\cite{liu2018stamp} effectively captures both user's general and current interests by utilizing memory priority components and short-term attention models.
    \item \textbf{CASER}\cite{tang2018personalized} employs convolutional operations from horizontal and vertical perspectives for personalized sequence recommendation.
    \item \textbf{SASRec}\cite{kang2018self} introduces a self-attention based sequence model to capture long-term semantics.
    \item \textbf{BERT4Rec}\cite{sun2019bert4rec} utilizes the Cloze task and bidirectional transformer modules to model user behavior in sequential recommendation. 
\end{itemize}

Moreover, we compare our proposed approach against five state-of-the-art denoising models. To make a fair comparison, we adopt the same base model (i.e., BERT4Rec) as previous works for our method. The details of all comparing models are as follows:
\begin{itemize}
    \item \textbf{DSAN}\cite{yuan2021dual} learns a trainable virtual target embeddings and employs adaptive sparse attention to filter irrelevant items.
    \item \textbf{FMLP-Rec}\cite{zhou2022filter} utilizes Fast Fourier Transform (FFT) and a low-pass filter to reduce noise impact for sequence representations. 
    \item \textbf{HSD+BERT4Rec}\cite{zhang2022hierarchical} eliminates noisy items from the sequence  by applying both sequence-level and user-level signals, enabling the learning of a refined and denoised sequence representation.
    \item \textbf{AC-BERT4Rec}\cite{zhou2023attention} improves the estimation and correction of item attention weights in historical interaction sequences through spatial and adversarial calibrators.
    \item \textbf{MSDCCL+BERT4Rec}\cite{zhu2024multilevel} captures user's long- and short-term interests by introducing target items and employs hybrid denoising strategies to reduce the influence of noise items on the recommendation model.
\end{itemize}

\begin{table*}[htbp]
\centering
% \caption{Evaluating the efficacy of diverse sequential recommendation strategies, with (w) or without (w/o) the MGSD-WSS, across a spectrum of five distinct datasets. The best score is in bold. All improvements are statistically significant (i.e., two-sided t-tests with $p < 0.05$).}
\caption{Results on five datasets comparing methods with MGSD-WSS (w) and without (w/o). All improvements are significant (two-sided t-test, $p<0.05$). Best results are in \textbf{bold}.}
\scalebox{0.9}{
    \begin{tabular}{llcccccccccccc}
    \toprule
    \multirow{2}{*}{Dataset}    &   \multirow{2}{*}{Metric}     &   \multicolumn{2}{c}{GRU4Rec} &  \multicolumn{2}{c}{NARM}    &   \multicolumn{2}{c}{STAMP}   &   \multicolumn{2}{c}{Caser}   &  \multicolumn{2}{c}{SASRec}  &   \multicolumn{2}{c}{BERT4Rec}    \\
    \cmidrule{3-14}
    &&   w/o &   w   &   w/o &   w   &   w/o &   w   &   w/o &   w   &   w/o &   w   &   w/o &   w  \\ \midrule
    \multirow{7}{*}{ML-100k}    
                                &   HR@5    &   0.0191  &   \textbf{0.0772} &   0.0180  &   \textbf{0.0704} &   0.0201  &   \textbf{0.0781} &   0.0212  &   \textbf{0.0623} &   0.0191  &   \textbf{0.0719} &   0.0191  &   \textbf{0.0716} \\
                                &   HR@10   &   0.0286  &   \textbf{0.1336} &   0.0403  &   \textbf{0.1198} &   0.0392  &   \textbf{0.1432} &   0.0339  &   \textbf{0.1128} &   0.0371  &   \textbf{0.1266} &   0.0414  &   \textbf{0.1360} \\
                                &   HR@20   &   0.0594  &   \textbf{0.2350} &   0.0657  &   \textbf{0.2163} &   0.0700  &   \textbf{0.2509} &   0.0679  &   \textbf{0.2017} &   0.0764  &   \textbf{0.2310} &   0.0912  &   \textbf{0.2439} \\
                                &   NDCG@5  &   0.0104  &   \textbf{0.0475} &   0.0132  &   \textbf{0.0415} &   0.0115  &   \textbf{0.0492} &   0.0113  &   \textbf{0.0363} &   0.0114  &   \textbf{0.0457} &   0.0117  &   \textbf{0.0470} \\
                                &   NDCG@10 &   0.0134  &   \textbf{0.0656} &   0.0202  &   \textbf{0.0572} &   0.0176  &   \textbf{0.0701} &   0.0153  &   \textbf{0.0525} &   0.0172  &   \textbf{0.0630} &   0.0189  &   \textbf{0.0662} \\
                                &   NDCG@20 &   0.0212  &   \textbf{0.0909} &   0.0267  &   \textbf{0.0814} &   0.0253  &   \textbf{0.0969} &   0.0238  &   \textbf{0.0748} &   0.0270  &   \textbf{0.0892} &   0.0315  &   \textbf{0.0932} \\
                                &   MRR@20  &   0.0109  &   \textbf{0.0520} &   0.0162  &   \textbf{0.0450} &   0.0132  &   \textbf{0.0555} &   0.0119  &   \textbf{0.0405} &   0.0139  &   \textbf{0.0512} &   0.0157  &   \textbf{0.0527} \\ \midrule
    \multirow{7}{*}{Beauty}    
                                &   HR@5    &   0.0077  &   \textbf{0.0464} &   0.0120  &   \textbf{0.0384} &   0.0080  &   \textbf{0.0380} &   0.0072  &   \textbf{0.0203} &   0.0242  &   \textbf{0.0470} &   0.0060  &   \textbf{0.0472} \\
                                &   HR@10   &   0.0135  &   \textbf{0.0725} &   0.0209  &   \textbf{0.0612} &   0.0135  &   \textbf{0.0622} &   0.0133  &   \textbf{0.0363} &   0.0386  &   \textbf{0.0733} &   0.0127  &   \textbf{0.0737} \\
                                &   HR@20   &   0.0256  &   \textbf{0.1075} &   0.0367  &   \textbf{0.0918} &   0.0231  &   \textbf{0.0955} &   0.0235  &   \textbf{0.0599} &   0.0561  &   \textbf{0.1077} &   0.0204  &   \textbf{0.1080} \\
                                &   NDCG@5  &   0.0045  &   \textbf{0.0303} &   0.0071  &   \textbf{0.0253} &   0.0046  &   \textbf{0.0239} &   0.0044  &   \textbf{0.0121} &   0.0129  &   \textbf{0.0307} &   0.0037  &   \textbf{0.0309} \\
                                &   NDCG@10 &   0.0064  &   \textbf{0.0387} &   0.0099  &   \textbf{0.0326} &   0.0064  &   \textbf{0.0317} &   0.0064  &   \textbf{0.0173} &   0.0175  &   \textbf{0.0392} &   0.0059  &   \textbf{0.0394} \\
                                &   NDCG@20 &   0.0094  &   \textbf{0.0475} &   0.0139  &   \textbf{0.0403} &   0.0088  &   \textbf{0.0401} &   0.0090  &   \textbf{0.0232} &   0.0219  &   \textbf{0.0478} &   0.0078  &   \textbf{0.0480} \\
                                &   MRR@20  &   0.0051  &   \textbf{0.0308} &   0.0077  &   \textbf{0.0261} &   0.0049  &   \textbf{0.0248} &   0.0051  &   \textbf{0.0132} &   0.0122  &   \textbf{0.0312} &   0.0044  &   \textbf{0.0314} \\ \midrule
    \multirow{7}{*}{Sports}    
                                &   HR@5    &   0.0064  &   \textbf{0.0180} &   0.0099  &   \textbf{0.0201} &   0.0071  &   \textbf{0.0261} &   0.0069  &   \textbf{0.0107} &   0.0113  &   \textbf{0.0266} &   0.0055  &   \textbf{0.0277} \\
                                &   HR@10   &   0.0114  &   \textbf{0.0299} &   0.0138  &   \textbf{0.0336} &   0.0123  &   \textbf{0.0413} &   0.0115  &   \textbf{0.0183} &   0.0175  &   \textbf{0.0428} &   0.0104  &   \textbf{0.0436} \\
                                &   HR@20   &   0.0183  &   \textbf{0.0465} &   0.0223  &   \textbf{0.0520} &   0.0182  &   \textbf{0.0634} &   0.0178  &   \textbf{0.0301} &   0.0268  &   \textbf{0.0658} &   0.0167  &   \textbf{0.0664} \\
                                &   NDCG@5  &   0.0035  &   \textbf{0.0116} &   0.0058  &   \textbf{0.0129} &   0.0046  &   \textbf{0.0170} &   0.0046  &   \textbf{0.0065} &   0.0059  &   \textbf{0.0172} &   0.0036  &   \textbf{0.0178} \\
                                &   NDCG@10 &   0.0051  &   \textbf{0.0154} &   0.0073  &   \textbf{0.0172} &   0.0062  &   \textbf{0.0218} &   0.0061  &   \textbf{0.0090} &   0.0079  &   \textbf{0.0224} &   0.0051  &   \textbf{0.0229} \\
                                &   NDCG@20 &   0.0068  &   \textbf{0.0196} &   0.0094  &   \textbf{0.0219} &   0.0077  &   \textbf{0.0274} &   0.0077  &   \textbf{0.0119} &   0.0102  &   \textbf{0.0282} &   0.0067  &   \textbf{0.0286} \\
                                &   MRR@20  &   0.0036  &   \textbf{0.0122} &   0.0059  &   \textbf{0.0136} &   0.0048  &   \textbf{0.0174} &   0.0049  &   \textbf{0.0070} &   0.0055  &   \textbf{0.0178} &   0.0040  &   \textbf{0.0182} \\ \midrule
    \multirow{7}{*}{Yelp}    
                                &   HR@5    &   0.0057  &   \textbf{0.0233} &   0.0113  &   \textbf{0.0313} &   0.0060  &   \textbf{0.0220} &   0.0045  &   \textbf{0.0273} &   0.0293 &   \textbf{0.0346}  &   0.0087  &   \textbf{0.0287} \\
                                &   HR@10   &   0.0102  &   \textbf{0.0401} &   0.0187  &   \textbf{0.0510} &   0.0099  &   \textbf{0.0368} &   0.0084  &   \textbf{0.0404} &   0.0352  &   \textbf{0.0550} &   0.0159  &   \textbf{0.0481} \\
                                &   HR@20   &   0.0184  &   \textbf{0.0675} &   0.0315  &   \textbf{0.0814} &   0.0161  &   \textbf{0.0608} &   0.0146  &   \textbf{0.0605} &   0.0439  &   \textbf{0.0859} &   0.0273  &   \textbf{0.0793} \\
                                &   NDCG@5  &   0.0034  &   \textbf{0.0145} &   0.0075  &   \textbf{0.0204} &   0.0038  &   \textbf{0.0141} &   0.0028  &   \textbf{0.0185} &   \textbf{0.0251} &   0.0230  &   0.0054  &   \textbf{0.0181} \\
                                &   NDCG@10 &   0.0048  &   \textbf{0.0199} &   0.0099  &   \textbf{0.0267} &   0.0051  &   \textbf{0.0188} &   0.0040  &   \textbf{0.0228} &   0.0270 &   \textbf{0.0295}  &   0.0077  &   \textbf{0.0243} \\
                                &   NDCG@20 &   0.0068  &   \textbf{0.0268} &   0.0131  &   \textbf{0.0343} &   0.0066  &   \textbf{0.0249} &   0.0055  &   \textbf{0.0278} &   0.0292 &   \textbf{0.0373}  &   0.0105  &   \textbf{0.0322} \\
                                &   MRR@20  &   0.0037  &   \textbf{0.0157} &   0.0081  &   \textbf{0.0214} &   0.0040  &   \textbf{0.0151} &   0.0031  &   \textbf{0.0188} &   \textbf{0.0250}  &  0.0239  &   0.0060  &   \textbf{0.0193} \\ \midrule
    \multirow{7}{*}{ML-1M}    
                                &   HR@5    &   0.0194  &   \textbf{0.1806} &   0.0151  &   \textbf{0.2131} &   0.0232  &   \textbf{0.1686} &   0.0104  &   \textbf{0.1643} &   0.0397  &   \textbf{0.1632} &   0.0224  &   \textbf{0.2066} \\
                                &   HR@10   &   0.0373  &   \textbf{0.2632} &   0.0349  &   \textbf{0.3056} &   0.0440  &   \textbf{0.2520} &   0.0215  &   \textbf{0.2532} &   0.0666  &   \textbf{0.2492} &   0.0495  &   \textbf{0.2958} \\
                                &   HR@20   &   0.0690  &   \textbf{0.3659} &   0.0591  &   \textbf{0.4101} &   0.0677  &   \textbf{0.3543} &   0.0589  &   \textbf{0.3569} &   0.1007  &   \textbf{0.3487} &   0.0980  &   \textbf{0.4005} \\
                                &   NDCG@5  &   0.0135  &   \textbf{0.1217} &   0.0080  &   \textbf{0.1446} &   0.0150  &   \textbf{0.1113} &   0.0063  &   \textbf{0.1069} &   0.0207  &   \textbf{0.1079} &   0.0132  &   \textbf{0.1405} \\
                                &   NDCG@10 &   0.0190  &   \textbf{0.1483} &   0.0144  &   \textbf{0.1745} &   0.0218  &   \textbf{0.1381} &   0.0099  &   \textbf{0.1356} &   0.0294  &   \textbf{0.1356} &   0.0218  &   \textbf{0.1693} \\
                                &   NDCG@20 &   0.0270  &   \textbf{0.1742} &   0.0205  &   \textbf{0.2009} &   0.0278  &   \textbf{0.1640} &   0.0194  &   \textbf{0.1617} &   0.0379  &   \textbf{0.1607} &   0.0339  &   \textbf{0.1957} \\
                                &   MRR@20  &   0.0159  &   \textbf{0.1204} &   0.0100  &   \textbf{0.1416} &   0.0168  &   \textbf{0.1105} &   0.0091  &   \textbf{0.1070} &   0.0203  &   \textbf{0.1080} &   0.0169  &   \textbf{0.1379} \\
    \bottomrule
    \label{tab:overall_mainstream_model}
    \end{tabular}
}
\end{table*}

\subsection{Implementation Details}
Following\cite{yuan2021dual, zhang2022hierarchical, zhu2024multilevel}, we use an embedding size of 100 and a mini-batch size of 256 for all models. The embedding parameters are initialized with a Gaussian distribution. We utilize the Adam optimizer with a default learning rate $10^{-3}$ and adopt an early-stopping training strategy on HR@20 metric. The regularization coefficient is determined through a search on $\beta$ in $\{0, 10^{-3}, 10^{-4}\}$. We utilize $10$ Gaussian kernels\cite{yang2023wsdms, wu2023adversarial, yang2022reinforcement}, one of which takes exact match (i.e., mean value $\mu_{0}=1.0$ and standard deviation $\sigma_{0}=10^{-3}$), the remaining kernels have $\sigma_{j}=0.1$ and their $\mu_{j}$ is evenly distributed within the range $\left[-1, 1\right]$. The initial temperature parameter is set to $\tau=0.5$ and decays after every 40 batches. For the baseline models, we either adhere to the original papers' performance or fine-tune and validate them for optimal performance.

%\zhu{20250318}

\begin{table*}[htbp]
\centering
% \caption{A comparative performance evaluation is conducted between MGSD-WSS enhanced base model (i.e., BERT4Rec) and the state-of-the-art denoising methodologies across five datasets. The best score and the second best score are in bold and underlined, respectively. Here * denotes statistically significant improvement (measured by a two-sided t-test with $p < 0.05$) over the best baseline. OOM: Out of memory on 24GB DGX.}
\caption{Performance of BERT4Rec with MGSD-WSS versus state-of-the-art denoising baselines on five datasets. Best and second-best scores are in \textbf{bold} and \underline{underlined}, respectively. * denotes statistically significant improvement over the best baseline (two-sided t-test, $p<0.05$). OOM: out of memory on 24GB DGX.}
\begin{tabular}{llccccccccc}
    \toprule
    Dataset &   \multicolumn{3}{l}{Model}   &   HR@5    &   HR@10   &   HR@20   &   NDCG@5  &   NDCG@10 &   NDCG@20 &   MRR@20  \\ \midrule
    \multirow{7}{*}{ML-100k}    
        &   \multicolumn{3}{l}{DSAN (AAAI'21)}           &   0.0201  &   0.0435   &   0.0700  &   0.0115  &   0.0188  &   0.0254  &   0.0133   \\
        &   \multicolumn{3}{l}{FMLP-Rec (WWW'22)}        &   0.0170  &   0.0477   &   0.0764  &   0.0117  &   0.0216  &   0.0288  &   0.0160   \\
        &   \multicolumn{3}{l}{HSD+BERT4Rec (CIKM'22)}   &   0.0339  &   0.0732   &   0.1294  &   0.0178  &   0.0305  &   0.0447  &   0.0218   \\
        &   \multicolumn{3}{l}{AC-BERT4Rec (CIKM'23)}    &   0.0212  &   0.0445   &   0.0877  &   0.0113  &   0.0188  &   0.0320  &   0.0147   \\
        &   \multicolumn{3}{l}{MSDCCL+BERT4Rec (2024)}   &   \underline{0.0718}  &   \underline{0.1270}   &   \underline{0.2126}  &   \underline{0.0449}  &   \underline{0.0631}  &   \underline{0.0845}  &   \underline{0.0498}   \\
        &   \multicolumn{3}{l}{MGSD-WSS+BERT4Rec}           &   \textbf{0.0761}*     &   \textbf{0.1360}* &   \textbf{0.2439}* &   \textbf{0.0470}* &   \textbf{0.0662}* &   \textbf{0.0932}* &   \textbf{0.0527}* \\ 
        \cmidrule{2-11}
        &   \multicolumn{3}{l}{Improv.}    &   5.99\% &   7.09\% &   14.72\% &   4.68\% &   4.91\% &   10.30\%  &   5.82\%   \\ \midrule
        					
    \multirow{7}{*}{Beauty}    
        &   \multicolumn{3}{l}{DSAN (AAAI'21)}           &   0.0092  &   0.0152   &   0.0264  &   0.0058  &   0.0077  &   0.0105  &   0.0062   \\
        &   \multicolumn{3}{l}{FMLP-Rec (WWW'22)}        &   0.0095  &   0.0166   &   0.0284  &   0.0056  &   0.0078  &   0.0107  &   0.0060   \\
        &   \multicolumn{3}{l}{HSD+BERT4Rec (CIKM'22)}   &   0.0261  &   0.0447  &   0.0683  &   0.0147  &   0.0207  &   0.0266  &   0.0151   \\
        &   \multicolumn{3}{l}{AC-BERT4Rec (CIKM'23)}    &   0.0200  &   0.0371  &   0.0609  &   0.0120  &   0.0175  &   0.0235  &   0.0133   \\
        &   \multicolumn{3}{l}{MSDCCL+BERT4Rec (2024)}   &   \textbf{0.0522}*  &   \underline{0.0714}  &   \underline{0.0955}  &   \textbf{0.0378}*  &   \textbf{0.0439}*  &   \textbf{0.0500}*  &   \textbf{0.0372}*   \\
        &   \multicolumn{3}{l}{MGSD-WSS+BERT4Rec}            &   \underline{0.0472}     &   \textbf{0.0737}* &   \textbf{0.1080}* &   \underline{0.0309} &   \underline{0.0394} &   \underline{0.0480} &   \underline{0.0314} \\ 
        \cmidrule{2-11}
        &   \multicolumn{3}{l}{Improv.}           &   -     &   3.22\% &   13.09\% &   - &   - &   - &   - \\ \midrule
        
    \multirow{7}{*}{Sports}    
        &   \multicolumn{3}{l}{DSAN (AAAI'21)}           &   0.0061  &   0.0105   &   0.0215  &   0.0042  &   0.0056  &   0.0084  &   0.0049   \\
        &   \multicolumn{3}{l}{FMLP-Rec (WWW'22)}        &   0.0068  &   0.0117   &   0.0180  &   0.0044  &   0.0059  &   0.0075  &   0.0046   \\
        &   \multicolumn{3}{l}{HSD+BERT4Rec (CIKM'22)}   &   0.0120  &   0.0190  &   0.0303  &   0.0078  &   \underline{0.0100}  &   0.0129  &   0.0081   \\
        &   \multicolumn{3}{l}{AC-BERT4Rec (CIKM'23)}   &   0.0112  &   0.0203  &   0.0351  &   0.0069  &   0.0099  &   0.0136  &   0.0077   \\
        &   \multicolumn{3}{l}{MSDCCL+BERT4Rec (2024)}   &   \underline{0.0271}  &   \underline{0.0387}  &   \underline{0.0553}  &   \textbf{0.0192}*  &   \textbf{0.0229}*  &   \underline{0.0271}  &   \textbf{0.0192}*   \\
        &   \multicolumn{3}{l}{MGSD-WSS+BERT4Rec}           &   \textbf{0.0277}*     &   \textbf{0.0436}* &   \textbf{0.0664}* &   \underline{0.0178} &   \textbf{0.0229}* &   \textbf{0.0286}* &   \underline{0.0182} \\ 
        \cmidrule{2-11}
        &   \multicolumn{3}{l}{Improv.}           &   2.21\%     &   12.66\% &   20.07\% &   - &   0.00\% &   5.54\% &   - \\ \midrule
        
    \multirow{7}{*}{Yelp}    
        &   \multicolumn{3}{l}{DSAN (AAAI'21)}           &   0.0269  &   0.0369   &   0.0541  &   0.0211  &   0.0242  &   0.0285  &   \underline{0.0216}   \\
        &   \multicolumn{3}{l}{FMLP-Rec (WWW'22)}        &   0.0203  &   0.0294   &   0.0436  &   0.0142  &   0.0171  &   0.0207  &   0.0144   \\
        &   \multicolumn{3}{l}{HSD+BERT4Rec (CIKM'22)}   &   \textbf{0.0292}*  &   0.0408  &   0.0593  &   \textbf{0.0223}*  &   \textbf{0.0260}*  &   0.0307  &   \textbf{0.0228}*   \\
        &   \multicolumn{3}{l}{AC-BERT4Rec (CIKM'23)}   &   0.0286  &   \underline{0.0445}  &   \underline{0.0700}  &   \underline{0.0194}  &   \underline{0.0245}  &   \underline{0.0309}  &   0.0202   \\
        &   \multicolumn{3}{l}{MSDCCL+BERT4Rec (2024)}   &   0.0243  &   0.0418  &   0.0695  &   0.0151  &   0.0207  &   0.0277  &   0.0163   \\
        &   \multicolumn{3}{l}{MGSD-WSS+BERT4Rec}           &   \underline{0.0287}     &   \textbf{0.0481}* &   \textbf{0.0793}* &   0.0181 &   0.0243 &   \textbf{0.0322}* &   0.0193 \\
        \cmidrule{2-11}
        &   \multicolumn{3}{l}{Improv.}           &   -     &   8.09\% &   13.29\% &   - &   - &   4.21\% &    \\ \midrule
        
    \multirow{7}{*}{ML-1M}    
        &   \multicolumn{3}{l}{DSAN (AAAI'21)}           &   0.0098  &   0.0336   &   0.0651  &   0.0048  &   0.0122  &   0.0200  &   0.0081   \\
        &   \multicolumn{3}{l}{FMLP-Rec (WWW'22)}        &   0.0210  &   0.0449   &   0.0707  &   0.0120  &   0.0199  &   0.0263  &   0.0142   \\
        &   \multicolumn{3}{l}{HSD+BERT4Rec (CIKM'22)}   &   0.0477  &   0.0886   &   0.1399  &   0.0297  &   0.0429  &   0.0558  &   0.0328   \\
        &   \multicolumn{3}{l}{AC-BERT4Rec (CIKM'23)}   &   OOM  &   OOM  &   OOM  &   OOM  &   OOM  &   OOM  &   OOM   \\
        &   \multicolumn{3}{l}{MSDCCL+BERT4Rec (2024)}   &  \underline{0.1351}     &   \underline{0.2095} &   \underline{0.3062} &   \underline{0.0873} &   \underline{0.1112} &   \underline{0.1356} &   \underline{0.0881} \\
        &   \multicolumn{3}{l}{MGSD-WSS+BERT4Rec}           &   \textbf{0.2066}*     &   \textbf{0.2958}* &   \textbf{0.4005}* &   \textbf{0.1405}* &   \textbf{0.1693}* &   \textbf{0.1957}* &   \textbf{0.1379}* \\ 
        \cmidrule{2-11}
        &   \multicolumn{3}{l}{Improv.}           &   52.92\%     &   41.19\% &   30.80\% &   60.94\% &   52.25\% &   44.32\% &   56.53\% \\ \midrule
    \bottomrule
    \label{tab:overall_denoising_model}
    \end{tabular}
\end{table*}

%, with the best results highlighted in bold and the second-best results underscored. 
\subsection{Main Performance Comparison}
To validate our model, we conduct two experiments on five datasets. We integrate the MGSD-WSS module into several state-of-the-art backbones and into BERT4Rec. Results for the state-of-the-art backbones are shown in Table \ref{tab:overall_mainstream_model}, and results for the BERT4Rec backbone are shown in Table \ref{tab:overall_denoising_model}.
The detailed results are presented below:

% We present the experimental results of various models across five datasets in Table \ref{tab:overall_mainstream_model} and Table \ref{tab:overall_denoising_model}. 
% We observe statistical significance for all of the improvements, which use a two-sided t-test compared to the baselines.
% Based on these experimental findings, we draw the following conclusions:
\begin{itemize}
    \item Table \ref{tab:overall_mainstream_model} shows the experimental results of various mainstream sequence recommendation models incorporating MGSD-WSS compared to the original baseline on all datasets. The integration of MGSD-WSS with baseline models results in a significant enhancement in performance, indicating the effectiveness of our proposed model in eliminating noisy items in historical interaction sequences.
    Notably, the proposed MGSD-WSS+BERT4Rec framework demonstrates consistent superiority over the original baseline across all datasets, with respective performance gains of 167.43\%, 195.87\%, and 235.67\% in terms of  HR@20, NDCG@20, and MRR@20 on the ML-100k dataset.
    %The performance of our proposed denoising model on the five data sets is better than their respective original baselines. 
    %Notably, the proposed MGSD-WSS framework demonstrates significant improvements over the original baseline on all datasets, and achieving respective enhancements of 167\%, 196\%, and 236\% across HR@20, NDCG@20, and MRR@20 evaluation metrics in ML-100k.

    %Since the original baseline model has different learning ability on different datasets, appropriate baseline models should be adopted for different datasets to achieve the best denoising effect.
    
    % \zhu{While most denoised models outperform their respective original baselines across the five datasets, we find that the optimal denoising approach varies among datasets due to differences in the learning capabilities of various mainstream sequence recommendation models on different data.} Additionally, we note that the denoised SASRec model on the Yelp dataset performs suboptimally in terms of NDCG@5 and MRR@20, due to the SASRec recommender being trained on the HR@20 metric, leading to an imbalance in the model's learning capabilities across different metrics. 
    \item Table \ref{tab:overall_denoising_model} compares the performance of our proposed denoising model MGSD-WSS with other denoising baselines. For a fair comparison, we employ MGSD-WSS combined with BERT4Rec in the experiment. We find that MGSD-WSS+BERT4Rec outperforms other denoising baselines across most metrics, particularly on ML-100k and ML-1M. 
    % This verifies that the generated noise supervised signal can significantly improve the ability of the model to recognize noisy items, and also confirms that the recognition of user noisy interests can enable the model to effectively learn the user's real intention.
    This shows that the generated noise supervised signals improve the model's ability to identify noisy items, while recognizing user's noisy interests helps model to learn their true intentions.
    % These results highlight the superiority and versatility of our approach. 
    %找一篇 有类似结果的论文参考
    %This confirms that, regardless of employing soft denoising or unsupervised hard denoising, the denoising process inherently remains incomplete and unreliable. These results highlight the superiority and versatility of our approach.
    However, on the Beauty, Sports, and Yelp datasets, MGSD-WSS+BERT4Rec performs suboptimally on minority metrics due to the relatively short average sequence lengths (i.e., 8.88 for Beauty, 8.32 for Sports, and 10.40 for Yelp). Introducing noise to short sequences contaminates the original sequence information, leading to less than optimal performance on certain metrics.
    % \zhu{However, on the Beauty, Sports, and Yelp datasets, MGSD-WSS+BERT4Rec performs suboptimally on minority metrics due to the relatively short average sequence lengths (i.e., 8.88 for Beauty, 8.32 for Sports, and 10.40 for Yelp). Introducing noise to short sequences contaminates the original sequence information, leading to less than optimal performance on certain metrics.}
\end{itemize}

\begin{table*}[tb]
\centering
% \caption{We present the performance of our model across five datasets in our ablation study, utilizing the metrics HR@20, NDCG@20, and MRR@20. The optimal results are emphasized in bold, while the second-best results are indicated with underlining.}
\caption{Ablation study performance across five datasets in terms of HR@20, NDCG@20, and MRR@20, where best results are highlighted in \textbf{bold}.}
\scalebox{1}{
    \begin{tabular}{llcccccc}
    \toprule
    Dataset &   Metric    &   Full   &   w/o MGP   &   w/o InSD  &   w/o ItSD &   w/o AND&   Improve.  \\ \midrule
    \multirow{3}{*}{ML-100k}
        &   HR@20   &   \textbf{0.2439} &   0.2331  &   0.2342  &   0.2115  &   0.2276  &   4.14\% \\
        &   NDCG@20 &   \textbf{0.0932} &   0.0892  &   0.0914  &   0.0821  &   0.0878  &   1.97\% \\
        &   MRR@20  &   \textbf{0.0527} &   0.0503  &   \textbf{0.0527}  &   0.0473  & 0.0500  &   0.00\% \\  \midrule
    \multirow{3}{*}{Beauty}
        &   HR@20   &   \textbf{0.1080} &   0.1065  &   0.1059  &   0.1059  &   0.1076  &   0.37\% \\
        &   NDCG@20 &   \textbf{0.0480} &   0.0468  &   0.0462  &   0.0472  &   0.0477  &   0.63\% \\
        &   MRR@20   &   \textbf{0.0314} &   0.0302  &   0.0297  &   0.0309  & 0.0311  &   0.96\% \\  \midrule
    \multirow{3}{*}{Sports}
        &   HR@20   &   \textbf{0.0664} &   0.0661  &   0.0583  &   0.0602 &   0.0654  &   0.45\% \\
        &   NDCG@20 &   \textbf{0.0286} &   0.0278  &   0.0241  &   0.0252  &   0.0279  &   2.51\% \\
        &   MRR@20   &   \textbf{0.0182} &   0.0173  &   0.0148  &   0.0157  & 0.0176  &   3.41\% \\  \midrule
    \multirow{3}{*}{Yelp}
        &   HR@20   &   \textbf{0.0793} &   0.0780  &   0.0773  &   0.0697  &   0.0748  &   1.67\% \\
        &   NDCG@20 &   \textbf{0.0322} &   0.0316  &   0.0316  &   0.0277  &   0.0303  &   1.90\% \\
        &   MRR@20   &   \textbf{0.0193} &   0.0190  &   0.0191  &   0.0163  & 0.0181  &   1.05\% \\  \midrule
    \multirow{3}{*}{ML-1M}
        &   HR@20   &   \textbf{0.4005} &   0.3624  &   0.3633  &   0.3782  &   0.3811  &   5.09\% \\
        &   NDCG@20 &   \textbf{0.1957} &   0.1706  &   0.1717  &   0.1807  &   0.1832  &   6.82\% \\
        &   MRR@20   &   \textbf{0.1379} &   0.1170  &   0.1172  &   0.1253  & 0.1276  &   8.07\% \\  
    \bottomrule
    \label{tab:ablation}
    \end{tabular}}
\end{table*}

\subsection{Ablation Study}
% Given that our proposed approach contains multiple modules, we conduct a series of experiments across five datasets to evaluate the effectiveness of each module. 
To thoroughly validate each module, we construct multiple model variants and evaluate their effectiveness on five datasets.
Specifically, we consider the following variants:
\begin{itemize}
    \item \textbf{w/o AND}: This variant removes the Auxiliary Noise Discrimination module from the model.
    \item \textbf{w/o InSD}: This variant eliminates the interest-granularity sequence denoising module from the model.
    \item \textbf{w/o ItSD}: This variant excludes the item-granularity sequence denoising module from the model.
    \item \textbf{w/o MGP}: This variant excludes the Multiple Gaussian-kernel Perceptron module from the model, relying solely on a Transformer encoder to process the historical interaction sequences.
    \item \textbf{Full}: This includes all the modules proposed in our approach.
\end{itemize}

\begin{table*}[htbp]
\centering
% \caption{In the interest- and item-wise, we employ the metrics HR@20, NDCG@20 and MRR@20 to illustrate the impact of weighted contrastive learning and traditional(i.e, without weights) contrastive learning on the performance of MGSD-WSS+BERT4Rec in ML-100k dataset. The optimal results are emphasized in bold.}
\caption{Performance comparison of weighted vs.\ traditional contrastive learning (CL) on the ML-100k dataset. 
% Item-wise and interest-wise indicate performance impacts at the item- and interest-granularity, respectively.
}
\scalebox{0.9}{
    \begin{tabular}{lccccccc}
    \toprule
    \multirow{2}{*}{Variants}   &   \multicolumn{2}{c}{Item-granularity CL} &  \multicolumn{2}{c}{Interest-granularity CL} & \multirow{2}{*}{HR@20} &   \multirow{2}{*}{NDCG@20} &   \multirow{2}{*}{MRR@20}    \\
    \cmidrule{2-5}
    &   Weighted &   Traditional   &   Weighted &   Traditional   &&&  \\ \midrule
    \multirow{1}{*}{Interest- and Item-granularity Traditional CL} &     &   \checkmark  &     &   \checkmark  &   0.2179   &   0.0847 &   0.0485 \\   \midrule
    \multirow{1}{*}{Interest-granularity Traditional CL} &   \checkmark  &     &     &   \checkmark  &   0.2243   &   0.0862 &   0.0488 \\
    \multirow{1}{*}{Item-granularity Traditional CL} &     &   \checkmark  &   \checkmark  &     &   \underline{0.2341}   &   \underline{0.0904} &   \underline{0.0514} \\   \midrule
    \multirow{1}{*}{Interest- and Item-granularity Weighted CL} &   \checkmark  &     &   \checkmark  &     &   \textbf{0.2450}   &   \textbf{0.0937} &   \textbf{0.0531} \\ 
    \multirow{1}{*}{Improv.} &     &     &     &     &   4.66\%   &   3.65\% &   3.31\% \\ \midrule	
    \bottomrule
    \label{tab:tranditional_CL_vs_Weighted_CL}
    \end{tabular}
}
\end{table*}

%Moreover, we observe that the interest- and item-granularity sequence denoising modules have the most significant impact on the performance of model in all datasets, highlighting the importance of denoising historical interaction sequences at different levels.

Table \ref{tab:ablation} shows the ablation study results. 
Specifically, removing any module leads to performance declines of varying magnitudes:
\begin{itemize}
\item Removing the Auxiliary Noise Discrimination module (w/o AND) causes the largest performance drop.
This is mainly because the module optimizes the model’s ability to distinguish noise at both the interest- and item-granularity during training. When it is removed, the model cannot effectively separate noise, resulting in a significant performance drop.

\item Removing the interest-granularity denoising module (w/o InSD) significantly degrades performance on the Beauty, Sports, and ML-1M datasets. Conversely, omitting the item-granularity denoising module (w/o ItSD) has the greatest impact on ML-100k and Yelp. These findings show that effective denoising must address both interest- and item-granularity noise, rather than focusing solely on item-granularity noise as prior work has done.
Omitting the Multiple Gaussian-kernel Perceptron module (w/o MGP) leads to a performance drop, indicating their effectiveness in suppressing noise.

\item Notably, on the ML-100k dataset, removing the interest-granularity denoising module (w/o InSD) leaves MRR unchanged but lowers both HR and NDCG. This module enhances the identification and ranking of secondary and long-tail items within the top-K recommendations, while having little effect on predicting the top-1 item. Without it, the model still retrieves the first relevant item but cannot secure additional hits or maintain overall ranking quality, causing HR and NDCG to drop even as MRR remains stable.
\end{itemize}

% As shown from the experimental results in Table \ref{tab:ablation}, the removal of any component leads to a decline in model performance. Notably, the exclusion of the Auxiliary Noise Discrimination module results in a drop in performance, thereby validating our motivation to enhance the denoising capability of the interest- and item-granularity sequence denoising modules by generating noise supervised signals through adding noise to the original historical interaction sequence. This approach effectively eliminates noisy items and noisy interests while preserving valuable information. 
% Moreover, we observe that on the Beauty, Sports, and ML-1M datasets, the interest-granularity sequence denoising module have the most significant impact on model performance. Similarly, on the ML-100k and Yelp datasets, the item-granularity sequence denoising module have the most substantial effect on model performance. These findings collectively underscore the importance of denoising the user's historical interaction sequence at different granularity levels.
% In addition, we observe the significance of the Gaussian kernel functions in enhancing the performance of the model.
%, concurrently substantiating the potential existence of noise within the original user history sequences.

\begin{figure*}[htb!]  %当前位置排版，放不下可以在本页顶部或者底部
      \centering  %居中
      \includegraphics[width=17cm,keepaspectratio]{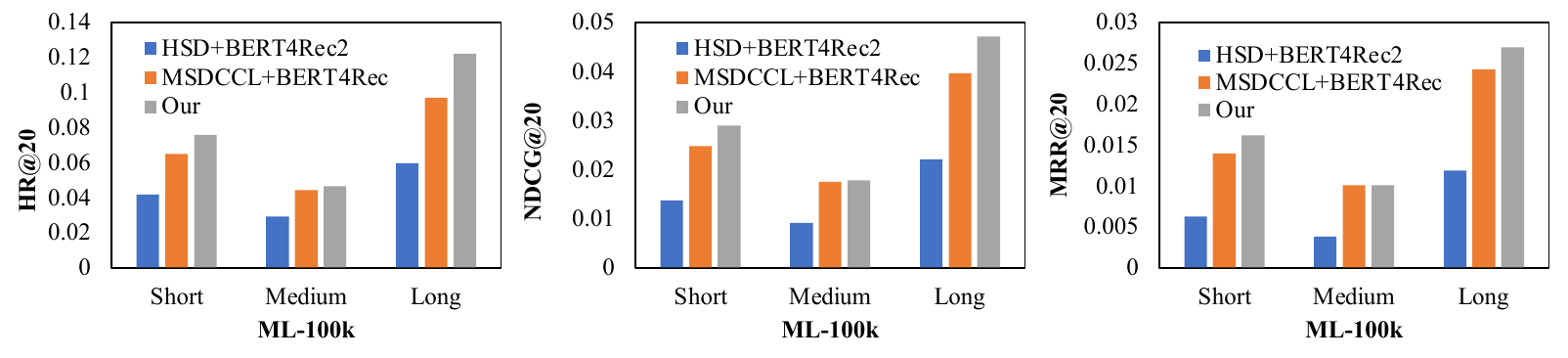} %photo是文件名，如果当前文件夹下没有同名不同格式的文件，比如photo.png就不需要后缀名啦
      \caption{Performance comparison of our model with two state-of-the-art denoising baselines (HSD+BERT4Rec and MSDCCL+BERT4Rec) across varying sequence lengths on the ML-100k dataset.} 
      \label{fig:different_seq_length} %指定给图片一个标签
\end{figure*}
\begin{figure}[!t]\centering
	\includegraphics[width=6cm]{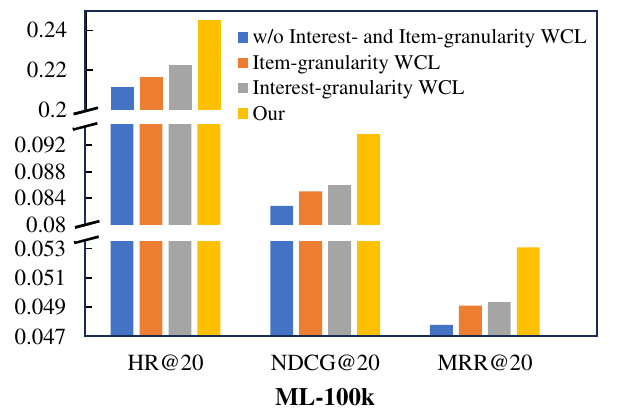}
	% \caption{The impact of using Interest- or Item-granularity weighted contrastive learning on model performance on the ML-100k dataset.}
    \caption{Impact of interest- and item-granularity weighted contrastive learning (WCL) on model performance in the ML-100k dataset.}
    \label{fig: weighted_CL}
\end{figure}

\subsection{Analysis of Weighted Contrastive Learning}
% \subsection{Comparison of Weighted and Traditional Contrastive Learning}
We evaluate weighted contrastive learning by comparing it with traditional (unweighted) contrastive learning at both the interest- and item-granularity.
Then, we ablate the interest- and item-granularity weighted contrastive learning to further analyze each module’s effect.
% We evaluate weighted contrastive learning by comparing it with traditional (unweighted) contrastive learning at both the interest and item levels.
\begin{itemize}
  \item Table \ref{tab:tranditional_CL_vs_Weighted_CL} shows the effect of weighted vs.\ traditional contrastive learning on model performance. Compared to traditional contrastive learning, weighted contrastive learning improves HR@20, NDCG@20, and MRR@20 by 12.59\%, 10.63\%, and 9.48\%, respectively. This indicates that assigning precise weights to guide the model away from noisy samples further enhances recommendation performance, a mechanism not supported by traditional contrastive learning.

  \item Figure \ref{fig: weighted_CL} compares the performance of interest- and item-granularity weighted contrastive learning. The results indicate that both modules significantly improved performance. This confirms the effectiveness of weighted contrastive learning. Moreover, interest-granularity weighted contrastive learning yields larger performance gains than item-granularity weighting. This advantage comes from guiding the model to avoid more noisy samples at the higher (interest) level, whereas item-granularity weighting only filters out a smaller set of noisy items at the lower (item) level.
\end{itemize}

\begin{figure*}[htb]  %当前位置排版，放不下可以在本页顶部或者底部
      \centering    \includegraphics[width=17cm,keepaspectratio]{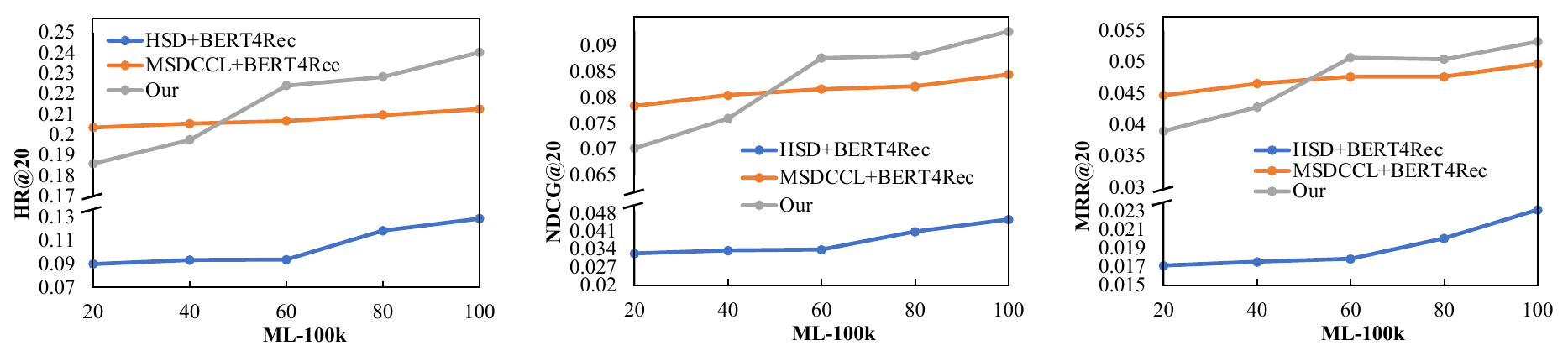}  
      % \caption{Performance of our model and the two best performing baseline (i.e., HSD+BERT4Rec and MSDCCL+BERT4Rec) with different training set proportion on  ML-100k.}  
      \caption{Performance comparison of our model and the strongest baselines (HSD+BERT4Rec, MSDCCL+BERT4Rec) on ML-100k under varying training data proportions.}
      \label{fig:different_train_set_ratio}  
\end{figure*}

\subsection{Comparison with Different Sequence Length}
% 用户的历史交互序列长度反映了用户交互行为的稀疏性。为了评估用不同序列长度对模型推荐性能的影响，我们按照交互序列将数据集(例如ML-100k)的长度等平频地分成三组(即短、中、长)。
% 我们将模型的性能与两个表现最好的去噪基线进行了比较，包括HSD+BERT4Rec和MSDCCL+BERT4Rec。
% 根据图*所示的结果，我们可以看到所有去噪模型在长序列上的性能提升最大，这是因为在长序列上噪音item在用户交互序列中所占比例更小，因而模型更够根据更加丰富的交互信息来生成高质量的降噪信号。
% 同时，我们还发现在中等长度上，模型的提升最小，这是可能是因为在中序列内噪音item所占比例更大，导致模型无法正确识别噪音item。
% 在所有三个组中，我们的模型始终优于HSD+BERT4Rec和MSDCCL+BERT4Rec。主要原因是我们通过显性地向序列内添加噪音，保证了噪音检测信号的质量，进而提升了模型的推荐能力。

% ml-100k的划分为short（0-28） 200/944=0.22，medium（29-49）182/944=0.19，long（50）562/944=0.60
\textbf{Experimental Setup}.
We compare our model with two strongest denoising baselines, i.e., HSD + BERT4Rec and MSDCCL + BERT4Rec, on the ML-100k dataset.
Specifically, we evaluate model performance on long, medium, and short sequences using HR@20, NDCG@20, and MRR@20. Sequences are grouped by interaction length: 
those with the highest number of interactions (e.g., 50 in ML-100k) are selected for the long-sequence group, and the remaining sequences are selected as evenly as possible into the medium- and short-sequence groups.
% Performance is measured by HR@20, NDCG@20 and MRR@20 for three uniformly defined sequence lengths: short (the first one-third of each user’s history sequences), medium (the first two-thirds) and long (the full history sequences).

\textbf{Experimental Results}.
As shown in Figure \ref{fig:different_seq_length}, all models achieve highest performance on long sequences. This is because the rich context, containing both true interactions and noise, enables effective signal-noise separation.
Performance on short sequences is moderate. This is because fewer interactions limit preference learning, while reduced noise makes denoising easier.
Models perform worst on medium-length sequences. This is because they still contain significant noise but lack the context needed for effective denoising, resulting in the greatest performance drop.

In all three scenarios, our model consistently outperforms the two strongest denoising baselines. This is mainly because we introduce labeled noise items that supply accurate noise signals, enabling the model to distinguish valid items from noise more effectively and achieve optimal performance. Furthermore, this consistent advantage demonstrates the robustness of our model.

\subsection{Robustness under Limited Data}
% 为了研究我们提出的模型在不同训练集比率下的有效性，我们将我们的模型的性能在数据集ML-100k上，与表现最好的基线HSD+BERT4Rec和MSDCCL+BERT4Rec进行了比较。
% 图*显示了当我们将训练集比例从20％变化到100％，步长为20％时的结果。
% 从结果可以看出，两种方法的性能都随着训练数据的增长而逐渐提高。
% 不过，HSD+BERT4Rec模型在所有训练集比率下的性能都是最差的，这证明高质量的噪音检测信号对模型性能提升的重要性。
% 另外，在低比率下我们的模型性能低于MSDCCL+BERT4Rec，随着比率的增加我们的模型性能逐渐高于MSDCCL+BERT4Rec。这是因为在训练数据较少时，向训练数据集内添加噪音无疑会进一步损坏数据的质量，而随着训练数据的增加，模型能够根据更加丰富的数据来建模噪音items的检测信号，这证明了我们的模型在实际应用中的有效性。

To assess robustness to training data size, we evaluate our model against HSD + BERT4Rec and MSDCCL + BERT4Rec on ML-100k with training set ratios from 20\% to 100\% (in 20\% increments), as shown in Figure \ref{fig:different_train_set_ratio}. All methods improve as the training set ratio increases, but HSD + BERT4Rec performs worst at every ratio, primarily because it struggles to accurately detect noise signals. At low ratios, our model trails MSDCCL + BERT4Rec; as the training set grows, it steadily closes the gap and eventually outperforms MSDCCL + BERT4Rec. 
This is primarily due to injected noise amplifying sparsity in scarce training data; once the training set exceeds roughly 45\% of the full dataset, this detrimental effect subsides and performance steadily improves, demonstrating the model’s robustness in low-data regimes.

% In order to evaluate the effectiveness of our proposed model across different training set ratios, we compare its performance on the ML-100k dataset with the top-performing baselines, namely HSD+BERT4Rec and MSDCCL+BERT4Rec.
% The results of altering the training set ratio from 20\% to 100\%, in increments of 20\%, are illustrated in Figure \ref{fig:different_train_set_ratio}.
% The results indicate that the performance of all methods gradually improves with the increase in training data.
% Nonetheless, the HSD+BERT4Rec model demonstrates the least performance across all training set ratios, thereby affirming the significance of high-quality noise detection signals in bolstering model performance.
% At lower ratios, our model’s performance falls below MSDCCL+BERT4Rec, but as the ratio increases, our model gradually outperforms MSDCCL+BERT4Rec. This is attributable to the fact that the insertion of noise into the training dataset when data is limited unquestionably further impairs the data quality. Yet, with the augmentation of training data, the model is capable of modeling noise item detection signals based on more abundant data, thereby validating the practical effectiveness of our model.

\begin{figure*}[htbp]  %当前位置排版，放不下可以在本页顶部或者底部
      \centering  %居中
      \includegraphics[width=17cm,keepaspectratio]{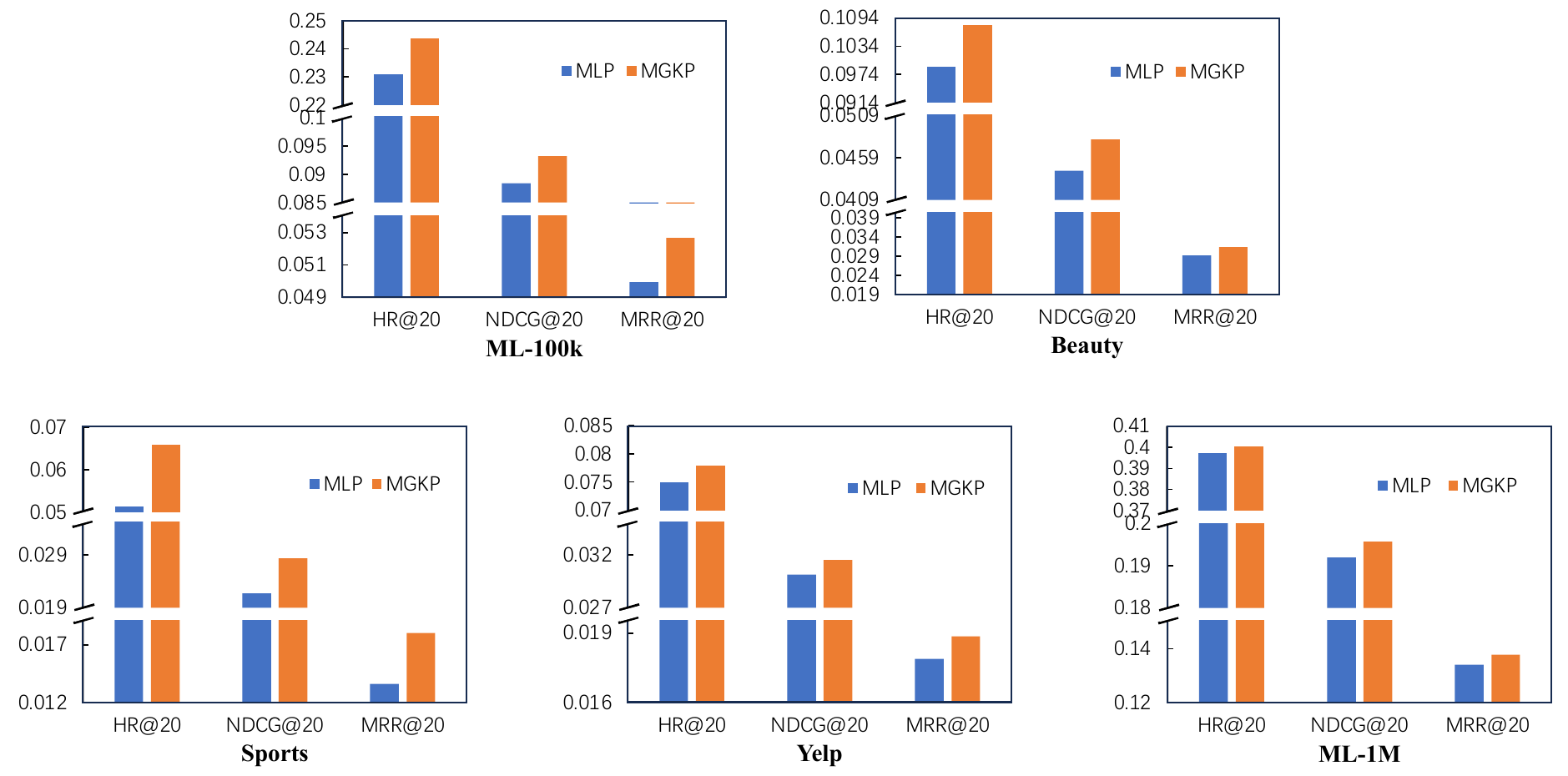} %photo是文件名，如果当前文件夹下没有同名不同格式的文件，比如photo.png就不需要后缀名啦
      % \caption{The impact of the Multiple Gaussian-kernel Perceptron (i.e., MGP) module and the Multilayer Perceptron (i.e., MLP) on the model’s performance across all datasets.} 
      \caption{Performance comparison between the variant model with Multilayer Perceptron (MLP) and the original model with Multiple Gaussian-kernel Perceptron (MGP).}
      \label{fig:gk_vs_mlp} %指定给图片一个标签
\end{figure*}

\subsection{Analysis of the Multiple Gaussian-kernel Perceptron}
% 为了验证Multiple Gaussian-kernel Perceptron模块在处理原始用户交互序列时，能够提升模型鲁棒性的能力。
% 我们将我们的模型在所有数据集上的性能指标，与使用多层感知机（即，MLP）替代Multiple Gaussian-kernel Perceptron的变体进行了比较。
% 图*展示了，在5个数据集上，Multiple Gaussian-kernel Perceptron模块对推荐带来的收益均高于使用多层感知机。
% 这是因为原始的用户历史交互序列含有噪音，需要利用target item和高斯核函数与先对原始输入进行增强，提高了模型的鲁棒性，而多层感知机只是对原序列进行空间变换。

To evaluate the impact of the Multiple Gaussian-kernel Perceptron, we replace it with a Multilayer Perceptron to create a variant model. We then compare the performance of this variant with the original model to assess the effect of the module. As shown in Figure \ref{fig:gk_vs_mlp}, the original model outperforms the variant model across multiple datasets. 
% This is mainly because the Gaussian kernel function maps noise into a higher-dimensional space, aiding in its separation and improving performance.
This is mainly because the Gaussian kernel function maps the both sequence into a common representation space, aiding in its effectively enhances the sequence representation of historical interactions.
In contrast, the Multilayer Perceptron only performs linear transformations on the original data, making it less effective.

\begin{figure*}[htbp]  %当前位置排版，放不下可以在本页顶部或者底部
      \centering  %居中
      \includegraphics[width=17cm,keepaspectratio]{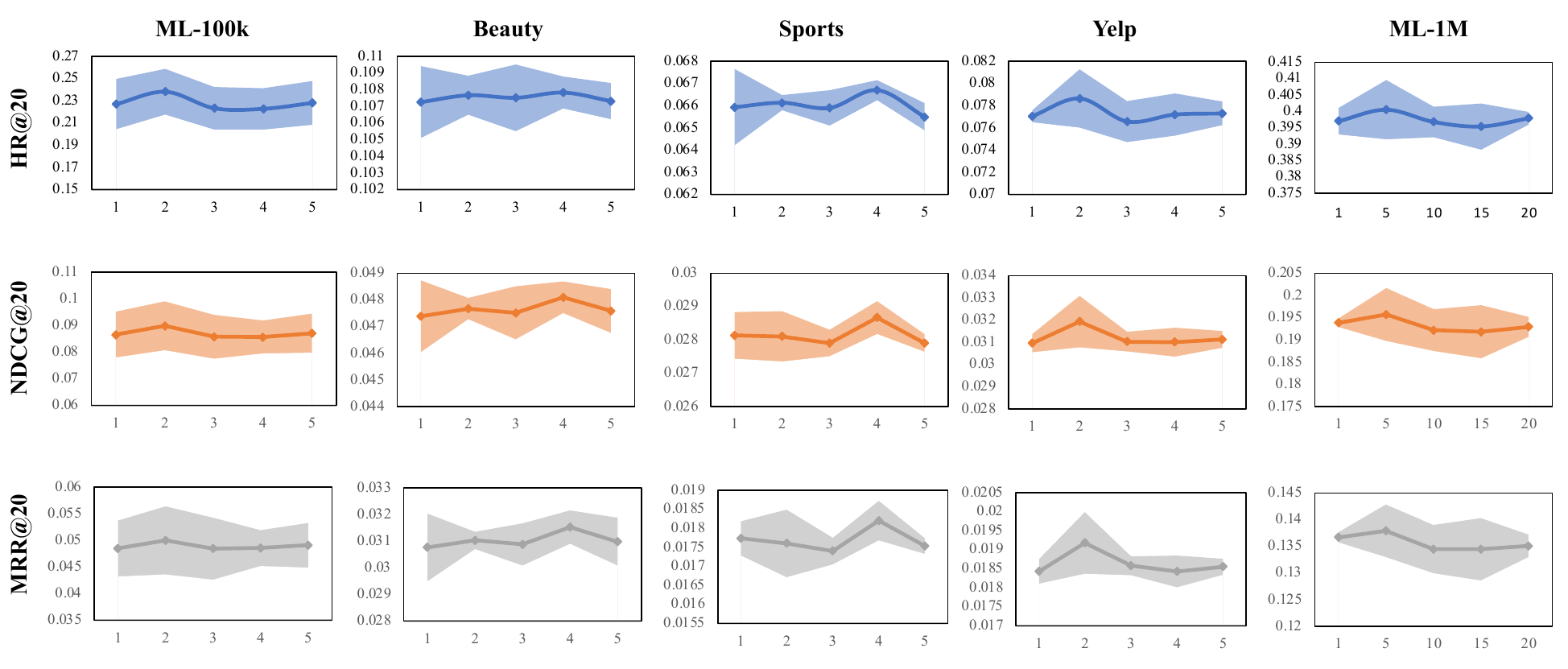} %photo是文件名，如果当前文件夹下没有同名不同格式的文件，比如photo.png就不需要后缀名啦
      \caption{Performance comparison with respect to the noise item number $t$ across five datasets.} 
      \label{fig:hyperparameter_sensitivity_noise_num} %指定给图片一个标签
\end{figure*}
\begin{figure*}[htbp]  %当前位置排版，放不下可以在本页顶部或者底部
      \centering  %居中
      \includegraphics[width=17cm,keepaspectratio]{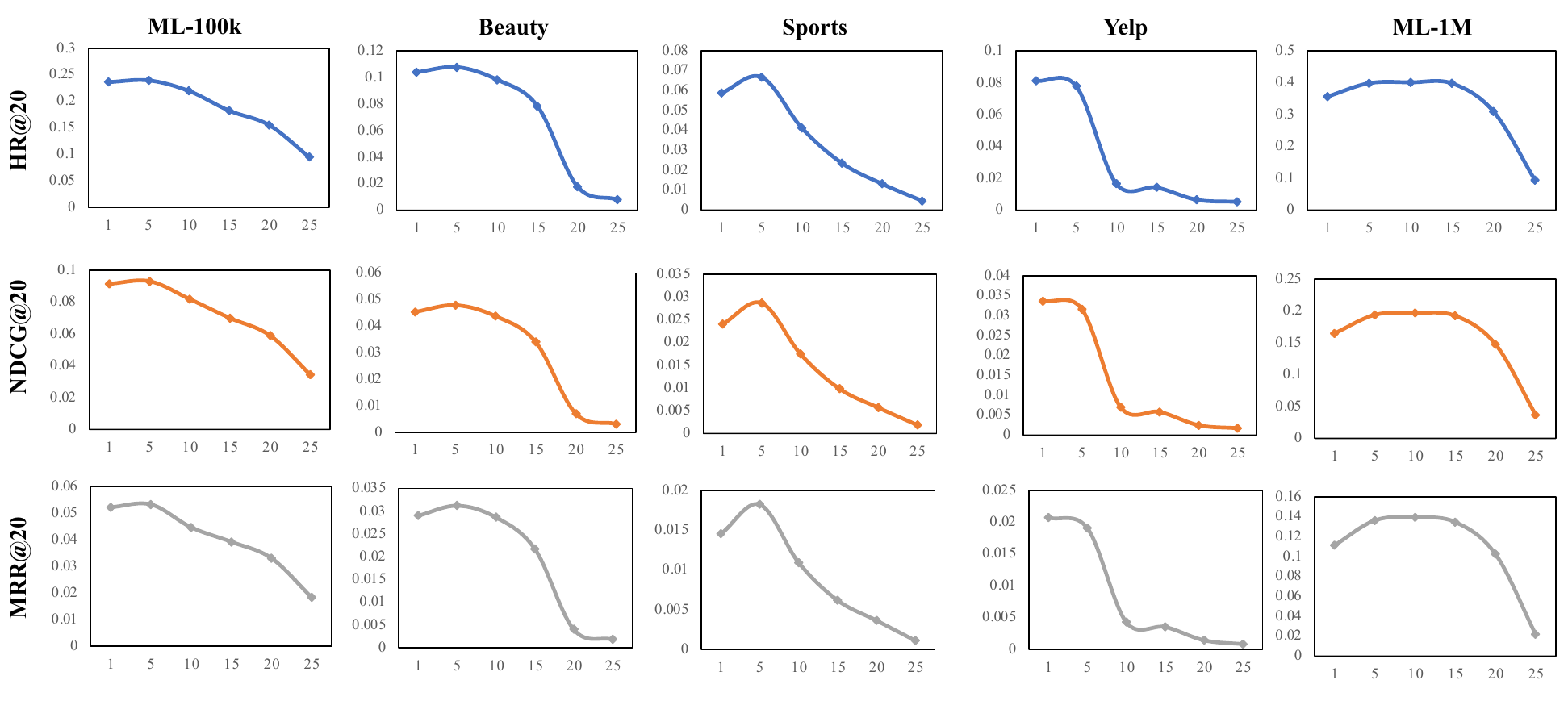} %photo是文件名，如果当前文件夹下没有同名不同格式的文件，比如photo.png就不需要后缀名啦
      \caption{Performance comparison with respect to the user interest number $m$ across five datasets.} 
      \label{fig:hyperparameter_sensitivity_interests_num} %指定给图片一个标签
\end{figure*}
\subsection{Hyperparameter Sensitivity Analysis}
% 在本节中，我们研究了超参数对MGSD-WSS+BERT4Rec模型推荐性能的影响。
% 我们首先探讨了向用户交互历史序列内添加噪音item的数量对模型性能的影响，它对生成高质量的噪音items检测信号至关重要。
% 由于用户的兴趣变化且多样的，因此我们还探索了不同的用户兴趣数量对推荐模型的性能影响。
% 为了保证实验可控，我们一次只改变一个超参数，同时固定其他超参数。

% In this section, we investigate the impact of hyperparameters on the recommendation performance of the MGSD-WSS+BERT4Rec model.
% We explore the impact of the number of noise items $t$ added to the user interaction history sequence on model performance, which is crucial for generating high-quality noise item detection signals.
% Given the variability and diversity of user interests, we further investigate the influence of different quantities of user interests $m$ on the performance of the recommendation model.
% To maintain experimental control, we alter a single hyperparameter at a time, concurrently keeping the rest constant.
In this section, we analyze the impact of two key parameters on model performance: the number of noise items $t$ and the number of user interests $m$.
Notably, when varying one parameter, the other is fixed at its optimal value to ensure fairness in the experiments.

\subsubsection{Effect of Noise Item Number $t$}
\textbf{Experimental Setup}.
In the experiments, we randomly select $t$ noisy items and insert them into the original sequences.
To avoid overly long sequences, we set a maximum sequence length based on the characteristics of each dataset: 200 for ML-1M and 50 for the others.
Sequences exceeding the maximum length are truncated accordingly.
To maintain a reasonable proportion of noise, we set $t \in [1, 5, 10, 15, 20]$ for ML-1M, and $t \in [1, 2, 3, 4, 5]$ for the other datasets.

\textbf{Experimental Results}.
As shown in Figure \ref{fig:hyperparameter_sensitivity_noise_num}, as $t$ increases, the performance first improves and then declines.
This is because a moderate number of labeled noise items helps the model learn to distinguish true preferences from noise by relying on accurate signals, leading to better performance.
However, when the number of noise items becomes too large, the performance starts to degrade. Excessive noise dilutes the informative signals in the sequence, making it harder for the model to capture accurate user preferences. Additionally, the model may overfit to noisy patterns, resulting in poorer overall performance.

\subsubsection{Effect of User Interest Number $m$}
% \textbf{Experimental Setup}.
We explore the impact of the number of user interests $m$ on model performance by varying $m \in [1, 5, 10, 15, 20, 25]$ across all datasets.

As shown in Figure \ref{fig:hyperparameter_sensitivity_interests_num}, performance initially improves with increasing $m$, peaking at $m$=5, and then begins to decline.
This suggests that a moderate number of user interests effectively capture true user preferences, leading to better performance. In contrast, too many interests may introduce excessive irrelevant information, thereby degrading performance.

Notably, the impact of user interests is most significant on the ML-1M and ML-100K datasets.
This is mainly because these datasets contain longer historical interaction sequences, which include more noise. 
An appropriate number of user interests helps the model extract meaningful preferences from noisy sequences, resulting in optimal performance.
This further demonstrates that user interests can effectively uncover true preferences even under noisy conditions, reinforcing the validity of the proposed module.

%%%%========================================================%%%%
\section{CONCLUSION}
In this paper, we proposed Multi-Granularity Sequence Denoising with Weakly Supervised Signal for Sequential Recommendation (MGSD-WSS) to address the challenges of noise in historical interaction sequences. The model first constructs weakly supervised noise signals to guide accurate noise identification. It then applies a hierarchical denoising approach, combining item-granularity and interest-granularity modules. The item-granularity denoising module utilizes noise-weighted contrastive learning to remove item-level noise, while the interest-granularity module refines user interest representations to address interest-level noise. Finally, based on the denoised sequence and interest representations, MGSD-WSS predicts the next item.
Extensive experiments on five benchmark datasets show that MGSD-WSS outperforms state-of-the-art sequence recommendation and denoising models.
% In future work, we will investigate advanced denoising methods that leverage weak supervision.
For future work, we will investigate the impact of different Gaussian kernel settings, e.g., adopting adaptive kernel parameters or fewer kernels. Moreover, we  attempt to investigate the influence of different noise generation techniques, such as adversarial or heuristic-based strategies. 
Finally, a theoretical or data-driven justification for the interest configuration is also needed to study.

%%%CONCLUSION end
%%%%=====================================================%%%%%%%%

%
% ---- Bibliography ----
%
% BibTeX users should specify bibliography style 'splncs04'.
% References will then be sorted and formatted in the correct style.
%

\printcredits

\section*{Acknowledgement}
This work was supported by the National Natural Science Foundation of China [grant number 62472059]; the Natural Science Foundation of Chongqing, China [grant number CSTB2022NSCQ-MSX1672]; the Chongqing Talent Plan Project, China [grant number CSTC2024YCJH-BGZXM0022]; the Major Project of Science and Technology Research Program of Chongqing Education Commission of China [grant number KJZD-M202201102].

% This work was supported by 

%% Loading bibliography style file
%\bibliographystyle{model1-num-names}
% \bibliographystyle{cas-model2-names}
\bibliographystyle{unsrt}
% Loading bibliography database
\bibliography{ref}
\end{document}